\makeatletter
\@namedef{ver@breakurl.sty}{9999/12/31}
\makeatother

\documentclass[a4paper,11pt]{article}
\def\Ab{{\boldsymbol A}}
\def\Bb{{\boldsymbol B}}

\def\Ib{{\boldsymbol I}}

\def\Pb{{\boldsymbol P}}

\def\Ub{{\boldsymbol U}}

\def\Xb{{\boldsymbol X}}

\def\mub{{\boldsymbol         \mu}}

\def\gb{{\boldsymbol g}}

\def\xb{{\boldsymbol x}}
\def\yb{{\boldsymbol y}}

\def\Sigmab{{\boldsymbol      \Sigma}}

\usepackage{amsmath}%AMS mathematical facilities for LATEX, When amsmath is loaded, AMS-LATEX packages amsbsy (for bold symbols), amsopn (for operator names) and amstext (for text embedded in mathematics) are also loaded.
\usepackage{amssymb} %provides an extended symbol collection. Will load amsfonts

\def\pre{u}
\def\vect{{\rm vec}}

\def\tilde{\widetilde}
\def\hat{\widehat}

\def\vec{{\rm vec}}

\def\df{{\rm df}}

\def\hat{\widehat}
\def\tilde{\widetilde}
\def\nn{\nonumber}

\usepackage{algorithmic}
\usepackage[ruled,linesnumbered,vlined]{algorithm2e} % An algorithm becomes a floating object (like figure, table, etc.)
\usepackage{mathptmx} % Use Times as default text font, and provide maths support, times+
\usepackage{verbatim} % LATEX verbatim
\usepackage{color} % Colour control for LATEX documents
\usepackage{url} % Verbatim with URL-sensitive line breaks
\usepackage{graphicx} % Enhanced support for graphics
\usepackage{array}% An extended implementation of the array and tabular environments which extends the options for column formats
\newcolumntype{?}{!{\vrule width 1pt}}
\usepackage{float}%Improves the interface for defining floating objects such as figures and tables
\usepackage{booktabs, threeparttable}%enhances the quality of tables 
\usepackage{subcaption} % loads the caption package
\usepackage{slashbox}
\usepackage{siunitx}
\usepackage{multirow} % Create tabular cells spanning multiple rows
\usepackage{tabularx} % Tabulars with adjustable-width columns
\usepackage{polynom} %  Macros for manipulating polynomials
\usepackage{threeparttable} % Tables with captions and notes all the same width
\usepackage{stackengine}
\RequirePackage[OT1]{fontenc}
\usepackage{csquotes}%Context sensitive quotation facilities
\usepackage{epstopdf,amsfonts,amsthm}
\usepackage{pdfpages}
\usepackage[dvipsnames]{xcolor}

\numberwithin{equation}{section}
\usepackage{titlesec}
\titleformat{\section}[block]{\Large\bfseries\raggedright}{\thesection}{1em}{}
\titleformat{\subsection}[block]{\bfseries}{\thesubsection}{3pt}{\space}
\titlespacing*{\subsection}{\parindent}{1ex}{1em}
\usepackage{enumitem}
\newcounter{daggerfootnote}

\allowdisplaybreaks
\graphicspath{{Figure/}}

\newcommand{\argmin}{\mathop{\rm argmin}}

\newcommand{\cE}{{\mathcal E}}

\newcommand{\R}{\mathbb{R}}

\newcommand{\VEC}{{\rm vec}}

\newcommand{\DIAG}{{\rm diag}}

\renewcommand{\hat}{\widehat}

\newcommand{\x}{\times}

\newtheorem*{theorem*}{Theorem}
\newtheorem*{corollary*}{Corollary}

\providecommand{\keywords}[1]
{
  \textbf{\textit{Keywords---}} #1
}
\usepackage{authblk}
\usepackage{hyperref}
\hypersetup{colorlinks,citecolor=blue,urlcolor=blue}

\begin{document}
%\renewcommand{\abstractname}{\vspace{-\baselineskip}}

%\begin{frontmatter}
%%\title{2SDR: Two Stage Dimension Reduction to Denoise Cryo-EM Images}
%\title[Two-stage dimension reduction]{Two-stage dimension reduction for noisy high-dimensional images and application to Cryogenic Electron Microscopy}
\title{Two-stage dimension reduction for noisy high-dimensional images and application to Cryogenic Electron Microscopy}

% 
%\begin{aug}
%\author{\fnms{Szu-Chi} \snm{Chung}\ead[label=e1]{phonchi@stat.sinica.edu.tw}},
%\address{Institute of Statistical Science\\
%         Academia Sinica\\
%         Tapei, Taiwan, 11529 \\
%         \printead{e1}}
%\author{\fnms{Shao-Hsuan} \snm{Wang}\ead[label=e2]{pico@stat.sinica.edu.tw }},
%\address{Institute of Statistical Science\\
%         Academia Sinica\\
%         Tapei, Taiwan, 11529 \\
%         \printead{e2}}
%\author{\fnms{Po-Yao} \snm{Niu}\ead[label=e3]{po.niu@email.ucr.edu}},
%\address{Institute of Statistical Science\\
%         Academia Sinica\\
%         Tapei, Taiwan, 11529 \\
%         \printead{e3}}
%\author{\fnms{Su-Yun} \snm{Huang}\ead[label=e4]{syhuang@stat.sinica.edu.tw}},
%\address{Institute of Statistical Science\\
%         Academia Sinica\\
%         Tapei, Taiwan, 11529 \\
%         \printead{e4}}
%\author{\fnms{Wei-Hau} \snm{Chang}\ead[label=e5]{weihau@chem.sinica.edu.tw}},
%\address{Institute of Chemistry\\
%         Academia Sinica\\
%         Tapei, Taiwan, 11529 \\
%         \printead{e5}}
%\author{\fnms{I-Ping} \snm{Tu}\thanksref{t2}\thanksref{t3}\ead[label=e6]{iping@stat.sinica.edu.t%w}}
%\address{Institute of Statistical Science\\
%         Academia Sinica\\
%         Tapei, Taiwan, 11529 \\
%         \printead{e6}}
%\thankstext{t2}{Corresponding author}
%\thankstext{t3}

%
%\affiliation{Institute of Statistical Science, Academia Sinica, Taiwan$^*$ \\
%Institute of Chemistry, Academia Sinica, Taiwan$^\dag$}

\author[1]{Szu-Chi Chung}
\author[1]{Shao-Hsuan Wang}
\author[1]{Po-Yao Niu}
\author[1]{Su-Yun Huang}
\author[2]{Wei-Hau Chang}
\author[1]{I-Ping Tu \footnote{Corresponding author} \footnote{The authors are grateful to the Editor-in-Chief, Tze Leung Lai, for comments which improved the contents and its presentation significantly.This work is supported by Academia Sinica:AS-GCS-108-08 and MOST:106-2118-M-001-001 -MY2.}}
\affil[1]{Institute of Statistical Science, Academia Sinica, Taiwan}
\affil[2]{Institute of Chemistry, Academia Sinica, Taiwan}

%\end{aug}
%\vskip .2in
\maketitle
%\vskip .2in

\begin{abstract}
Principal component analysis (PCA) is arguably the most widely used dimension-reduction method for vector-type data. When applied to a sample of images, PCA requires vectorization of the image data, which in turn
entails solving an eigenvalue problem for the sample covariance matrix.
We propose herein a two-stage  dimension reduction (2SDR) method  for image reconstruction from high-dimensional noisy image data. The first stage treats the image as a matrix, which is a tensor of order 2, and uses multilinear principal component  analysis (MPCA) for matrix rank reduction and image denoising. The second stage vectorizes the reduced-rank matrix and achieves further dimension and noise reduction. Simulation studies demonstrate excellent performance of 2SDR, for which we also develop an asymptotic theory that establishes consistency of its rank selection. Applications to cryo-EM (cryogenic electronic microscopy), which has revolutionized structural biology, organic and medical chemistry, cellular and molecular physiology in the past decade, are also provided and illustrated with benchmark cryo-EM datasets. Connections to other contemporaneous developments in image reconstruction and high-dimensional statistical inference are also discussed.

%based on a  statistical  model with  two layers of noise structure.
%2SDR first uses multi-linear PCA (MPCA) \cite{BioHung2012, GLRAM1} to extract core scores from the images and  screen the first layer of noise, and then uses PCA %on these scores to  reduce the second layer of noise.
%This approach inherits the advantage from MPCA in computation and PCA in guaranteeing uncorrelated projection scores.
%Testing with several cryo-electron microscopy (cryo-EM) benchmark particle datasets shows that 2SDR performs better in denoising than MPCA and PCA alone. 
%Further application of 2SDR to mixture of 3D structures followed by tSNE
%visualization shows it is capable of disentangling similar molecular
%conformations, thereby demonstrating its usage in high dimension data
%beyond 2D images.
\end{abstract}
%\keywords{Dimension Reduction, Image Analysis, Kronecker Product, cryo-Electron Microscopy,
%Rank Selection}

%\begin{keyword}[class=AMS]
%\kwd[Primary ]{00K00}
%\kwd{00K01}
%\kwd[; secondary ]{00K02}
%\end{keyword}

%%  Upper case for every keyword
%\begin{keyword}
%\kwd{Generalized Information Criterion}
%\kwd{Image Denoising and Reconstruction}
%\kwd{Random Matrix Theory}
%\kwd{Rank Selection}
%\kwd{Stein's Unbiased Estimate of Risk}
%\end{keyword}
\keywords{Generalized Information Criterion, Image Denoising and Reconstruction, Random Matrix Theory, Rank Selection, Stein's Unbiased Estimate of Risk.}

%\end{frontmatter}
%\newpage
%\end{document}

%%%%%%%%%%%%%%%%%%%%%%%%%%%%%%%%
%%         Section 1          %% 
%%%%%%%%%%%%%%%%%%%%%%%%%%%%%%%%
\section{Introduction}\label{c:intro}
As noted by Chen et al. \cite{AASChen2014}, determining the 3D atomic structure of biomolecules is important for elucidating the physicochemical mechanisms underlying vital processes, and major breakthroughs in this direction led to Nobel Prizes in Chemistry awarded to Roger Kornberg in 2006, Venkatraman Ramakrishnan, Thomas Steitz and Ada Yonath in 2009, and Brian Kobilka and Robert Lefkowitz in 2012. While X-ray crystallography played an important role in the last two discoveries, most large proteins resisted attempts at crystallization. Cryogenic electron microscopy (cyro-EM), which does not need crystals and is therefore amenable to structural determination of proteins that are refractory to crystallization \cite{frank2006three}, has emerged as an alternative to X-ray crystallography for determining 3D structures of macromolecules in the past decade, culminating in the Nobel Prize in chemistry awarded to Jacques Dubochet, Joachim Frank and Richard Henderson in 2017. 

Chen et al. \cite[pp. 260-261]{AASChen2014} first describe the workflow of cryo-EM image analysis and then focus on  2D clustering step that identifies structurally homogeneous sets of images after these images have gone through the alignment and other image processing steps of the workflow. The size of a cryo-EM image is often larger than 100 pixels measured in each direction. Treating an image as a vector with dimension $p$, which is the pixel number that exceeds $100 \times 100 = 10^4$, clustering a large set of these high-dimensional vectors is a challenging task, particularly because of the low signal-to-noise ratio (SNR) in each cryo-EM image. Chen et al. \cite{AASChen2014} proposed to use a novel clustering method called $\gamma$-SUP, in which $\gamma$ refers to $\gamma$-divergence and SUP refers to "self-updating process", to address these challenges.  Section 1.1 gives an overview of $\gamma$-SUP, while Section 1.2 describes visualization of multidimensional data using t-distributed Stochastic Neighbor Embedding (t-SNE) plots \cite{maaten2008visualizing}. Section 1.3 provides an overview of multilinear principal component analysis (MPCA) which we developed its statistical properties in \cite{BioHung2012} and which will be used in Section 2 to develop a new two-stage dimension reduction (2SDR) method for high-dimensional noisy images. The first stage of 2SDR uses MPCA of a random matrix X $\in \mathbb{R}^{p \times q}$ to represent the image, together with consistent selection of the actual rank $(p_0,q_0)$ of the matrix $X$ as a second-order tensor; representing a high-dimensional image as a matrix has computational advantages over image vectorization. The second stage of 2SDR carries out PCA for the vectorized reduced-rank image to achieve further dimension reduction; Sections 1.4 and 1.5 give an overview of the literature on rank selection for MPCA and PCA. 
In Section 3,  applications of 2SDR to cryo-EM images are illustrated with  benchmark datasets including 70S ribosome and  80S ribosome.  We show that 2SDR can improve 2D image clustering to curate the clean particles and 3D classification to separate various conformations. In particular, for a dataset containing 11,000 volumes of 5 conformations of 70S ribosome structure, we demonstrate that the t-SNE plot of 2SDR shows clear separation of the 5 groups, which illustrates the promise  of the method for cryo-EM image analysis.  The 80S ribosome dataset, which has huge sample size and large pixel numbers, exemplifies the computational capability of 2SDR.
Section 4 gives further discussion and concluding remarks. %Section 3 shows how 25DR can be combined with t-SNE (t-distributed Stochastic Neighbor Embedding) \cite{maaten2008visualizing} for visualizing denoised (via 2SDR) cryo-EM images, or with $\gamma$-SUP for 2D clustering of these images. Further discussion and concluding remarks are given in Section 4.

\subsection{$\gamma$-SUP and 2D clustering of cryo-EM images}
An important step in the workflow of cryo-EM image analysis is 2D clustering to identify structurally and orientationally homogeneous sets of particle images. Sorzano et al \cite{Sorzano} proposed a k-means algorithm which, for a given number of clusters, iteratively bisects the data to achieve this via a kernel-based entropy measure
to mitigate the impact of outliers. Yang et al. \cite{Yang} subsequently noted the difficulty to find good
initial values and prespecify a manageable number of clusters, and found the k-means approach to be unsatisfactory. This led Chen et al. \cite{AASChen2014} to develop an alternative 2D clustering method called $\gamma$-SUP, in which $\gamma$ stands for ''$\gamma$-divergence'' and SUP is abbreviation
for ''self-updating process'' (introduced by Shiu and Chen \cite{Shiu}) and to demonstrate its superior performance in 2D clustering of cryo-EM images. Their $\gamma$-SUP algorithm can be viewed as an
iteratively reweighted procedure using kernel weights that are inversely proportional to the $\gamma$-divergence $D_{\gamma}(f||g)$ of the chosen working parametric model (with a density function $g$) from the empirical measure with a kernel density function $f$; $0<\gamma<1$ and the case $\gamma\rightarrow 0$ gives the Kullback-Leibler  divergence \cite{fujisawa2008robust}.
Chen et al. \cite{AASChen2014} propose to choose the working model of $ q$-Gaussian distribution \cite{Amari} so that the chosen model $g$ has the smallest  $\gamma$-divergence from $f$.
The $q$-Gaussian distribution over $\mathbb{R}^p$, with $q<1+2p^{-1}$, is a generalization of the Gaussian distribution (corresponding to $q\rightarrow 1$) and has a density function of the form
\begin{equation}
g_q(\xb; \mu,\sigma) = (\sqrt{2\pi\sigma})^{-p}{c_{p,q}} \, 
	\exp_q\left(-\|\xb-\mu\|^2/(2\sigma^2)\right),~~~ \xb\in{\mathbb R}^p, \nn
\end{equation}	
	where $\exp_q(u)$ is the $q-$exponential function $\exp_q(u) = \left\{1 + (1 - q)u \right\}_+^{\frac{1}{1 - q}}$ { and } $x_+ = \max(x, 0)$.
	For $1<q<1+2p^{-1}$, it corresponds to the multivariate $t$-distribution with $\nu=2(q-1)^{-1}-p$ degrees of freedom, whereas it has compact support for $q<1$, which is assumed by Chen et al. \cite{AASChen2014}
in applications to cryo-EM images and for which $c_{p,q}=(1-q)^{p/2}\Gamma(1+p/2+(1-q)^{-1})/\Gamma(1+(1-q)^{-1})$.

Instead of working with a  mixture density that requires specification of the number of mixture component  and therefore encounters the same difficulty as the $k$-means approach, Chen et al. \cite{AASChen2014} propose to fit each component $j$  with $g_q(\cdot;\mu_j,\sigma)$ separately but with same $\sigma$. For given $\sigma$, the minimizer $\mu_j^*$ of $D(f  \parallel  g_q(\cdot;\mu_j,\sigma))$ is given by the solution of the equation $\mu_j =\int \yb w(\yb; \mu_j,\sigma)dF(\yb)/
\int w(\yb; \mu_j,\sigma)dF(\yb)$, where $F$ is the  distribution function with density $f$. Hence replacing $F$ by the empirical distribution $\widehat{F}$ of the sample $\{ \yb_i, 1\leq i\leq n\}$ leads to the recursion
\begin{align*}
\widehat\mub^{(\ell+1)}_j =\frac{\int \yb w(\yb; \mub^{(\ell)}_j,\sigma)d\widehat F(\yb)}{
\int w(\yb; \mu^{(\ell)}_j,\sigma)d\widehat F(\yb)},~~\ell=0,1,\cdots
\end{align*}
Using  the SUP algorithm of \cite{Shiu} to replace $\widehat{F}$ by the empirical distribution $\widehat F^{(j)}$ of $\{\mu_i^{(j)}:~1\leq i \leq n \}$ leads to 
the $\gamma$-SUP recursion 
\begin{align*}
\widehat\mub^{(\ell+1)}_j =\frac{\int \yb w(\yb; \mub^{(\ell)}_j,\sigma)d\widehat F^{(\ell)}(\yb)}{
\int w(\yb; \mub^{(\ell)}_j,\sigma)d\widehat F^{(\ell)}(\yb)},~~\ell=0,1,\cdots,
\end{align*}
in which $w^{(\ell)}_{ij}$ has the explicit formula 
$
w_{ij}^{(\ell)}  =\exp_{1-s}\left(-\left\|(\widehat\mub_j^{(\ell)}-\widehat\mub_i^{(\ell)}) /\tau\right\|^2 \right)
$, where $\tau =\sqrt{2}\sigma/\sqrt{\gamma-(1-q)}>0$ and $s=(1-q)/\{\gamma -(1-q)\}>0$. This explicit formula is derived by Chen et al. \cite[p269]{AASChen2014} who also show that 
eventually ``
$\gamma$-SUP converges to certain
$K$ clusters, where $K$
depends on the tuning parameters ($\tau, s$) but otherwise is data-driven.'' Another advantage of $\gamma$-SUP is that $\sigma$ is absorbed in the tuning parameter $\tau$, hence selection of $\tau$ obviates the need to select $\sigma$. 
As pointed out by Chen et al. \cite[p.268]{AASChen2014}, $\gamma$-SUP "involves $(s,\tau)$ as the tuning parameters" and "numerical studies have found that $\gamma$-SUP is quite insensitive to the choice of $s$ and that $\tau$ plays the decisive role in the performance of $\gamma$-SUP", for which they suggest to use a small positive value (e.g. 0.025) of $s$ and a "phase transition plot" for practical implementation in their Section 4, where performance of a clustering method is measured by "purity" and "c-impurity" numbers that will be discussed below in Section 3.1.
%In Section \ref{c:sim}, we discuss the choice of $\tau$ and $s$ and dimension-reduction and computational schemes to implement $\gamma$-SUP efficiently.

\subsection{t-distributed stochastic neighbor embedding (t-SNE)}
\label{sec:tSNE}

Data visualization is the graphical representation of data, for which tools from multiple disciplines have been developed, including computer graphics, infographics, and statistical graphics. Recent advances, which include t-SNE and Laplacian eigenmap, focus on complex data belonging to a low-dimensional manifold embedded in a high-dimensional space. 
Laplacian eigenmap builds a graph from neighborhood information of the dataset. Using the adjacency matrix $w_{ij}$ to incorporate this neighborhood information, Belkin and Niyogi \cite{Laplacian} consider the optimization problem of choosing configuration points $y_i$ to minimize $\sum_{j>i}w_{ij}||y_j-y_i||^2$, which forces $y_i$ and  $y_j$ to be close to each othe if $w_{ij}$ is large, and apply graph theory and the  Laplace-Beltrami operator to formulate the optimization problem as an eigenvalue problem \cite{Chung2000}. However, until Maaten and Hinton \cite{maaten2008visualizing} developed t-SNE for data visualization in 2008, no visualization was able to separate the MNIST benchmark dataset, consisting of $28 \times 28$ pixel images each of which has a hand-written digit from 0 to 9, into ten groups (corresponding to the ten digits).

Similar to Laplacian eigenmap, t-SNE also has underpinnings in local information of the dataset $\{X_1,\dots,X_n\}$, with distance measure $d(X_i, X_j)$ between $X_i$ and $X_j$. Instead of using the adjacency  matrix of a graph to incorporate the local information, t-SNE uses a Gaussian Kernel to transform the distance matrix $d(X_i, X_j)$ first into a probability transition matrix $\tilde \pi_{ij} \propto \exp(-d(X_i, X_j)/2\sigma_i^2)$, i.e., $\tilde \pi_{ij}$ is the conditional probability of moving from position $X_j$ given the initial position $X_i$, and then into a probability 
mass function $p_{ij} = (\tilde \pi_{ij} + \tilde \pi_{ji})/(2n)$ over pairs of configuration points $y_i$ and $y_j$.  This probability distribution enables Maaten and Hinton \cite{maaten2008visualizing} to combine ideas from multidimensional scaling \cite{Mead} which minimizes $\sum_{j>i} (d_{ij}-d^*_{ij})^2$, where $d_{ij}$ (respectively, $d^*_{ij}$) is the Euclidean distance between configuration points $y_i$ and $y_j$ in the dataset (respectively, in a low-dimensional subspace of the high-dimensional space), to choose the $t$-distribution for "stochastic neighbor embedding" (hence t-SNE) that minimizes the Kullback-Leibler divergence $I(p,q) = \sum_{i} \sum_{j} p_{ij} \log (p_{ij}/q_{ij})$, where $q_{ij} =(1+||y_i-y_j||^2)^{-1}/\sum_{k\neq l}{(1+||y_k-y_l||^2)^{-1}}$ for configuration points $y_i$ and $y_j$ with $i \neq j$, which corresponds to the density function of the t-distribution for $||y_i-y_j||$ with one degree of freedom. Applying stochastic gradient to minimize $I(p,q)$ yields 
\begin{equation*}\label{eq:gradientYi}
\frac{\delta I}{\delta y_i} = 4 \sum_{j} (p_{ij}-q_{ij})(y_j-y_i)(1+||y_j-y_i||^2)^{-1},
\end{equation*}
which then leads to the iterative scheme
\begin{equation*}\label{eq:gradientupdate}
y_i^{(t)} = y_i^{(t-1)} + \eta (\frac{\delta I}{\delta y_i}) + \alpha^{(t)} (y_i^{(t-1)}-y_i^{(t-2)}),
\end{equation*}
where $\eta$ is the "learning rate" and $\alpha^{(t)}$ is the "momentum" at iteration $t$ of machine learning algorithms. How to choose them and $\sigma_i^2$ in $p_{ij}$ (via $\tilde \pi_{ij}$) is also discussed in  \cite{maaten2008visualizing}.

\subsection{PCA and MPCA}\label{MPCA_PCA}

Let $X, X_1,\cdots,X_n$ be i.i.d. $p \times q$ random matrices. Let  $y=\vec(X_i)$, where $\vec$ is the operator of matrix vectorization by stacking the matrix into a vector by columns. The statistical model for PCA is 
%%%%%%%%%%%%%%%%%%%%%%%%%%%%%%%%
%%         Eq 1.1            %%
%%%%%%%%%%%%%%%%%%%%%%%%%%%%%%%%
\begin{equation}\label{eq:PCA_vec}
 y=\mu + \Gamma\nu+\varepsilon, 
\end{equation}
\noindent
where $\mu$ is the mean, $\,\nu\in\R^r$  with $r \leq pq$, $\Gamma$ is a $pq \times r$ matrix with orthonormal columns, and  $\,\varepsilon$ is independent of $\nu$ with ${\rm E}(\varepsilon)=0$ and ${\rm Cov}(\varepsilon) = c\,I_{pq}$. 
The zero-mean vector $\nu$  has covariance matrix $\Delta = \DIAG (\delta_1,\delta_2,\cdots,\delta_r)\ $ with $\delta_1\ge\delta_2\ge\cdots \ge\delta_r>0\,$.
The estimate $\hat\Gamma$ contains the first $r$ eigenvectors of the sample covariance matrix $S_n=n^{-1}\sum_{i=1}^n(y_i-\overline y)(y_i-\overline y)^\top$, and $\vec(\overline X)+\hat \Gamma \hat \nu_i$ provides a reconstruction of the noisy data $\vec(X_i)$. The computational cost, which increases with both the sample size $n$ and the dimension $pq$, becomes overwhelming for high-dimensional data. For example, the 80S ribosome dataset in \cite{wong2014cryo} has more than $n=100,000$ images of dimension $pq$ (after vectorization) with $p=q=360$. The computational complexity of solving for the first $r$ eigenvectors of $S_n$ is $O((pq)^2r)=O(10^{10}\times r)$ in this case, as shown by Pan and Chen \cite{Pan1999}, which may be excessive for many users. An alternative to matrix vectorization is MPCA \cite{BioHung2012,GLRAM1} or higher-order singular value decomposition (HOSVD) \cite{SIAMLath2000b}, and both methods have been found to reconstruct images from noisy data reasonably well while MPCA has better asymptotic performance than HOSVD \cite{BioHung2012}, hence we only consider MPCA in the sequel. 

%and the images can be reconstructed as
%\begin{eqnarray}\label{eq:PCA_vec}
%\vec(\hat X_i)=\vec(\overline X)+\hat\Gamma \hat \nu_i
%=\vec(\overline X)+P_{\hat\Gamma}\vec(X_i-\overline X),
%\end{eqnarray}
%where $P_{\hat\Gamma}=\hat\Gamma\hat\Gamma^\top$ also called the projection matrix of $\hat\Gamma$ and $\overline X$ is the estimated mean. 

MPCA models the $p \times q$ random matrix $X$ as
%%%%%%%%%%%%%%%%%%%%%%%%%%%%%%%%
%%         Eq 1.2            %%
%%%%%%%%%%%%%%%%%%%%%%%%%%%%%%%%
\begin{equation}\label{eq:MPCA} 
\begin{aligned}
X &= Z +\cE  \in\R^{p\x q}, \quad Z = M +AUB^\top,
\end{aligned}
\end{equation}
where $M\in\R^{p\times q}$ is the mean, $U\in\R^{p_0\x q_0}$ is a random matrix with $p_0 \leq p,\,q_0 \leq q$, $A$ and $B$ are non-random $p \times p_0, q \times q_0$ matrices with orthogonal column vectors, $\cE$ is a zero-mean radnom vector independent of $U$ such that ${\rm Cov}(\VEC(\cE)) = \sigma^2\,I_{pq}$. Ye \cite{GLRAM1} proposed  to use generalized low-rank approximations of matrices to estimate $ A$ and $B$. Given $(p_0,q_0)$, $\hat A$ consists of the leading $p_0$ eigenvectors of the covariance matrix $\sum_{i=1}^n (X_i-\overline X) P_{\hat B} (X_i-\overline X)^\top$, and  $\hat B$ consists of the leading $q_0$ eigenvectors  of ${\sum_{i=1}^n
(X_i-\overline X)^\top P_{\hat A}  (X_i-\overline X)}$, where the matrix
$P_{\hat A}={\hat A}{\hat A}^\top$ (respectively, $P_{\hat B}={\hat B}{\hat
B}^\top$) is the projection operator into the span of the column vectors of $\hat A$ (respectively, $\hat B$). The estimates can be computed by an iterative procedure that usually takes no more than 10 iterations to converge.
%From the optimization formula, MPCA is to search $\hat A$ and $\hat B$ for all the data matrix 
%\begin{equation}
%\label{eq:min_sample}(\hat A, {\hat U_1, \ldots, \hat U_n}, \hat B)
%= \underset{\substack{A \in {\cal O}^{p\times  p_0} ,
%U_k \in \mathbb{R}^{ p_0 \times q_0} ,
%B \in \mathcal{O}^{q\times  q_0}}}
%{\argmin} \
%\frac{1}{n} \sum_{k=1}^n \| X_k -\overline X- A U_k B^\top \|_F^2.
%\end{equation}
%As for comparison, singular value decomposition (SVD) is to solve the eigenvectors for both column and row spaces on one data matrix say $X_1$ while MPCA is on a set of data matrix $\{X_1,\dots,X_n\}$.
%The optimization equation for SVD is as follows and the solution of $\hat A$ and $\hat B$ guarantees that $\hat D=\hat A^\top X_1\hat B$ to be a diagonal matrix while $\hat U_1,\dots,\hat U_n$ of MPCA not.
%\begin{equation}
%\label{eq:min_SVD}
%(\hat A, {\hat D}, \hat B)
%= \underset{\substack{A \in {\cal O}^{p\times  p_0} ,
%D \in \mathbb{R}^{ p_0 \times q_0} ,
%B \in \mathcal{O}^{q\times  q_0}}}
%{\argmin} \
 %\| X_1 - A D B^\top \|_F^2.
%\end{equation}
Replacing $A$ and $B$ by their estimates $\hat A$ and $\hat B$ in \eqref{eq:MPCA} yields 
%%%%%%%%%%%%%%%%%%%%%%%%%%%%%%%%
%%         Eq 1.3             %%
%%%%%%%%%%%%%%%%%%%%%%%%%%%%%%%%
\begin{align}\label{eq:MPCA_vec}
&& \hat U_i = \hat A^\top  (X_i-\overline X) \hat B, \ \mbox{hence } \hat A\hat U_i {\hat B^\top}= P_{\hat A}(X_i-\overline X)P_{\hat B}, 
%&&  \vec(\hat X_i)= \vec(\overline X)+P_{\hat B\otimes \hat A}\vec(X_i-\overline X), \nn
\end{align}
i.e., $\vec(\hat A\hat U_i {\hat B^\top}) = {P_{\hat B\otimes \hat A}}\vec(X_i-\overline X)$ where $\otimes$ denotes the Kronecker product and $P_{\hat B\otimes \hat A}=(\hat B\hat B^\top)\otimes (\hat A\hat A^\top)$. %is the projection matrix. 
%Hence $\vec(\hat X_i)= \vec(\overline X)+P_{\hat B\otimes \hat A}\vec(X_i-\overline X)$.

%The useful references include~\cite{BioHung2012,GLRAM1,SIAMLath2000b,SIAMLath2000a,SIAMRKolda2008}. 
%For MPCA we aim to solve for the minimizer  $(\hat A, \{\hat U_i\}_{i=1}^n, \hat B)$ defined as

%\begin{equation}
%\label{eq:min_sample}
%(\hat A, \{\hat U_i\}_{i=1}^n, \hat B)
%= \underset{\substack{A \in {\cal O}^{p\times p_0} \\
%U_i \in \mathbb{R}^{ p_0 \times q_0} \\
%B \in \mathcal{O}^{q\times q_0}}}
%{\argmin} \
%\frac{1}{n} \sum_{i=1}^n \| X_i -\overline X- A U_i B^\top \|_F^2 \, .
%\end{equation}

%For the observations  $\{y_i, ~1\le i\le n\}$, we have $y_i=\Gamma\nu_i+\varepsilon_i$.
%(\ref{eq:MPCA_vec}) implies that $\vec(\hat X_i-\overline X)$ is reconstructed by the bases $(\hat B\otimes\hat A)$ with the scores $\vec(U_i)=(\hat B\otimes\hat A)^\top \vec{(X_i-\overline X)}$. 
Hung et al. \cite[p.571]{BioHung2012} used the notion of the Kronecker envelope introduced by Li et al. \cite{li2010dimension} to connect MPCA and PCA models. For the PCA model ~\eqref{eq:PCA_vec} with $y=\vec(X)$, they note form Theorem 1 of \cite{li2010dimension} that there exists a full rank $p_0q_0 \times r$ matrix $G$ such that $\Gamma=( B\otimes  A)G$ for which ~\eqref{eq:PCA_vec} becomes 
%Schott \cite{schott2014tests} introduced the Kronecker envelope PCA (KEPCA) to connect the MPCA and PCA models. For the case of \eqref{eq:PCA_vec} and \eqref{eq:MPCA} considered herein, we take {\color{red}$\Gamma=( B\otimes  A)G$} for which  \eqref{eq:PCA_vec} has the form
%%%%%%%%%%%%%%%%%%%%%%%%%%%%%%%%
%%         Eq 1.4             %%
%%%%%%%%%%%%%%%%%%%%%%%%%%%%%%%%
\begin{equation}\label{eq:PCA}
\vec(X)= \VEC(M) + ( B\otimes  A)G\nu+\vec(\cE) \in \R^{pq}, \ \mbox{with} \  \nu \in \R^r.
\end{equation}
%Replacing $A$ and $B$ by $\hat A$ and $\hat B$ in the special case $\Gamma=(B\otimes A)G$ yields %vector form of the KEPCA model
%When taking $\Gamma=(B\otimes A)G$, the PCA model becomes the
%vector form of KEPCA model:
%\begin{equation}\label{eq:KEPCA}
%\vec(X_i)= \VEC(\overline X) +(B\otimes A)G\nu_i+\vec(\cE_i).
%\end{equation}
%for $1\le i\le n$. 
The components of $\nu$ in \eqref{eq:PCA} are pairwise uncorrelated, whereas those of $\vec(U)$ in the MPCA model \eqref{eq:MPCA} are not; the subspace ${\rm span}(B \otimes A)$ is "the unique minimal subspace that contains ${\rm span}(\Gamma)$" and is "called the Kronecker envelope of $\Gamma$". From \eqref{eq:MPCA_vec} and \eqref{eq:PCA}, we need to specify {$(p_0, q_0)$ and $r$}, respectively, and the following two subsections will summarize previous works on how to select them.

%Note that $v_i$ and $v_j$ are uncorrelated for $i \neq j$ in the sample version of (\ref{eq:KEPCA}) whereas the pairwise correlation of the $\hat U_i$ may be markedly different from 0. 

%Note that the random variable $\nu$ contains $r$ variables and 
%we refer the $r$-dimensional vector $\hat \nu_i$ as the projection scores for $X_i$ where $1\le i\le n$. The pairwise sample correlations among $r$ variables of $\{\hat \nu_i, 1\le i\le n\}$ are all zero, i.e., these $r$ variables are uncorrelated, while entries in $\{\hat U_i, 1\le i\le n\}$ are not.
%This suggests following up MPCA with KEPCA for dimension reduction.

%Comparing the reconstruction representation of PCA and MPCA by their vector forms %(\ref{eq:PCA}) and (\ref{eq:MPCA_vec}), we observe that PCA employs $\hat \Gamma$ as basis %to reconstruct the data and MPCA employs $(\hat B\otimes\hat A)$. Note that the pairwise %sample correlations among $r$ variables of $\{\hat \nu_i, 1\le i\le n\}$ are all zero, %i.e., these $r$ variables are uncorrelated, while entries in $\{\hat U_i, 1\le i\le n\}$ %are not. In our experience on many real datasets, the pairwise sample correlations among %variables of $\{\hat U_i, 1\le i\le n\}$  usually are highly correlated. This suggests %that $\{\hat U_i, 1\le i\le n\}$ can be further decorrelated and the dimension be reduced.

\subsection{Rank selection for MPCA} 
We first review the recent work of Tu et al. \cite{Iping2019} using Stein's unbiased risk estimate (SURE) to derive a rank selection method for the MPCA model, under the assumption that $\VEC(\cE)$ has a multivariate normal distribution with mean 0 and covariance matrix 
$\sigma^2I_{pq}$. They note that "with the advent of massive data, often endowed with tensor structures," such as in images and videos, gene-gene environment interactions, dimension reduction for tensor data has become an important but challenging problem. In particular, Tao et al. \cite{tao2007probabilistic} introduced a mode-decoupled probabilistic model for tensor data, assuming independent Gaussian random variables for the stochastic component of the tensor, and applied the information criteria in AIC and BIC independently to each projected tensor mode. In the special case of the MPCA model for tensors of order 2, the stochastic components correspond to the entries of $U$ in (\ref{eq:MPCA}), which are pairwise correlated and therefore not independent. For the MPCA model \eqref{eq:MPCA}, Tu et al. \cite{Iping2019} introduce the risk function
%shows that in order to decompose the tensor structure
%to independent multiple modes, the covariance of $\vec(U)$ needs to follow  a Kronecker product form
%as $\Sigma_{\vec(U)}=\Sigma_r\otimes\Sigma_c$ where $\Sigma_r$ and $\Sigma_c$ refer to the covariance matrix for the rwo and column spaces.
%%%%%%%%%%%%%%%%%%%%%%%%%%%%%%%%
%%         Eq 1.5             %%
%%%%%%%%%%%%%%%%%%%%%%%%%%%%%%%%
%where $I_m$ denotes the $m \times m$ identity matrix.
%With the model $X_k=Z_k+{\cE_k}$, the risk function is defined as
\begin{align}
    \label{eq:risk_ori}
        R(p_0, q_0; \sigma^2) & = \sum_{k=1}^n {\rm E}\| Z_k- \hat Z_k \|_F^2 \\
        & = \sum_{k=1}^n \Big [ {\rm E}\| X_k- \hat Z_k \| _F ^2 -
  2 {\rm E} [\mathrm{tr}\{ \cE_k^\top (X_k -\hat Z_k)\} ] +
  {\rm E}\|\cE_k\|_F^2\Big ] \nn
\end{align}
to be used as the risk that SURE estimates by using Stein's identity \cite{Stein, ulfarsson2008dimension} under the assumption that $\vec(\cE) \sim N(0, \sigma^2I_{pq})$. The last (i.e., third) summand in \eqref{eq:risk_ori} is $npq\sigma^2$, the first summand can be estimated by $\sum_{k=1}^n| X_k- \hat A \hat U_k \hat B^\top \|_F^2$, and the second summand is equal to 
%The first term in (\ref{eq:risk_ori}) can be estimated by the residual sum of squares and the third term is 
%$nm\sigma^2$.
%Thanks to the iid Normal assumption of $\cE$, we can apply Stein's identity \cite{Stein, ulfarsson2008dimension} to derive the second term as
%%%%%%%%%%%%%%%%%%%%%%%%%%%%%%%%
%%         Eq 1.6            %%
%%%%%%%%%%%%%%%%%%%%%%%%%%%%%%%%
\begin{align}\label{eq:second_term}
%- 2 {\rm E} [\mathrm{tr} \{ \cE_k^\top (X_k - \hat Z_k)\} ]
-2 \sigma^2\sum_{k=1}^n {\rm E} \left[ \mathrm{tr}\left\{ \frac{\partial  \vec (X_k -\hat Z_k)}
{\partial  \vec (X_k)^\top}\right\} \right] = -2npq\sigma^2 +  2 \sigma^2 \sum_{k=1}^n {\rm E}\left[ \mathrm{tr}
 \left\{ \frac{\partial  \vec (\hat Z_k)}{\partial  \vec (X_k)^\top}\right\} \right],
\end{align}
by Stein's identity. {Letting $\Sigma$ denote the covariance matrix of $\vec(X)$ in \eqref{eq:MPCA} and  $\hat\Sigma=n^{-1}\sum_{i=1}^n\vec(X_i-\bar X)\vec(X_i-\bar X)^\top$ be its estimate, Tu et al. \cite[pp. 35-37]{Iping2019} note that $\vec(\hat Z_k)=P_{\hat B\otimes \hat A} \vec(X_k)$ and that $\hat A$ and $\hat B$ depend on $\hat \Sigma$, and use the chain rule to compute the derivative  $(\frac{\partial}{\partial \vec (X_k)^\top})(P_{\hat B \otimes \hat A} \vec(X_k))=(\frac{\partial (P_{\hat B \otimes \hat A} \vec(X_k))}{\partial (\vec ({\hat\Sigma}))^\top})
(\frac{\partial \vec({\hat \Sigma})}{\partial \vec (X_k)^\top})$ } and then prove that {the second summand of} \eqref{eq:second_term} can be estimated by  $2\sigma^2 \df_{(p_0,q_0)}$, where
%where $m=pq$. Since $\hat Z_k=\hat A\hat A^\top X_k\hat B\hat B_k^\top$ and $\hat A$ and $\hat B$ all depend on the covariance matrix, thus the derivative $\frac{\partial  \vec (\hat Z_k)}{\partial 
%   \vec (X_k)^\top}$ can apply the chain rule through $\hat\Sigma=1/n\sum_{i=1}^n(\vec(X_i-\bar X))(\vec(X_i-\bar X))^\top$. The partial derivative of the eigenvector with respect to the covariance matrix is by the perturbation method on a set of stationary equations. The calculation details can be found in 
%   the appendix of \cite{Iping2019}.
%We then define the GDF, or simply the degrees of freedom, of the MPCA model as
%\sum_{k=1}^n   {\rm E}\left[ \mathrm{tr}\left\{ \frac{\partial  \vec (\hat Z_k)}{\partial  \vec (X_k)^\top}\right\} \right]=
%%%%%%%%%%%%%%%%%%%%%%%%%%%%%%%%
%%         Eq 1.7             %%
%%%%%%%%%%%%%%%%%%%%%%%%%%%%%%%%
\begin{align}\label{mpca_df}
\df_{(p_0, q_0)} = pq + (n - 1) p_0 q_0 +
     \sum_{i=1}^{p_0} \sum_{\ell= p_0+1}^p
     \frac{\hat \lambda_i + \hat \lambda_\ell}{\hat \lambda_i - \hat \lambda_\ell} +
     \sum_{j=1}^{q_0} \sum_{\ell= q_0+ 1}^q
     \frac{\hat \xi_j+\hat \xi_\ell }{\hat \xi_j - \hat \xi_\ell}
\end{align} 
is the "degree of freedom" and the  $\hat \lambda_i$ (respectively, $\hat \xi_j$)  are the eigenvalues, in decreasing order of their magnitudes, of $ n^{-1}\sum_{k=1}^n (X_k -\overline X) P_{\hat B} (X_k-\overline X)^\top$ (respectively, $ n^{-1}\sum_{k=1}^n (X_k -\overline X) P_{\hat A} (X_k-\overline X)^\top$). Their Section 3.2 gives interpretations and discussion of ${\rm df}_{(p_0, q_0)}$, and their proof of \eqref{mpca_df} keeps the numerator terms to be column vectors and denominator terms to be row vectors, and uses the identity $\vec(ABC)=(C^\top \otimes A) \vec(B)$ together with 
\begin{flalign*}
\sum_{i=1}^{n} X_i\hat B\hat B^\top X_i^\top
&=\sum_{i=1}^{n}X_i\left(\sum_{j=1}^{q_0}\hat b_j\hat b_j^\top\right) X_i^\top=\sum_{i=1}^{n}\sum_{j=1}^{q_0}(X_i\hat b_j)(X_i\hat b_j)^\top \\ &=\sum_{i=1}^{n}\sum_{j=1}^{q_0}\left((b_j^\top\otimes I_p)\vec(X_i)\right)\left(\vec(X_i)^\top(b_j\otimes I_p)\right) \\
&=n\sum_{j=1}^{q_0}\left\{(b_j^\top\otimes I_p)\hat\Sigma(b_j\otimes I_p)\right\}
\end{flalign*}
in which $\hat B = \{\hat b_1,\dots,\hat b_{q_0}\}$, where $\hat b_j$ is the normalized eigenvector of $n^{-1}\sum_{i=1}^nX_i^\top \hat A\hat A^\top X_i$ associated with the eigenvalue $\hat\xi_j$, and we have assumed $\overline X=0$ to simplify the notation involving $X_i-\overline X$. %The $\hat \lambda_i$ (respectively, $\hat \xi_j$) in \eqref{mpca_df} are the eigenvalues, in descending order of their magnitudes, of $ n^{-1}\sum_{k=1}^n (X_k
%-\overline X) P_{\hat B} (X_k-\overline X)^\top$ (respectively, with $P_{\hat B}$ replaced by $P_{\hat A}$). 

%where  the $\hat \lambda_i$s and $\hat \xi_j$s are the eigenvalues of $ \sum_{k=1}^n (X_k
%-\overline X) P_{\hat B} (X_k-\overline X)^\top/n$ and ${
%\sum_{k=1}^n (X_k-\overline X)^\top P_{\hat A} (X_k-\overline X)}/n$
%in descending order. 

%The first term of $\df_{(\tilde p,\tilde q)}$ reflects the degrees of freedom of the mean $\hat M =\overline X$ whose size is $m=p\times q$. The second term, $(n-1) \tilde p \tilde q$,
%reflects the degrees of freedom of the core matrices $U_1, \dots, U_n$, each of size $\tilde p\times \tilde q$. Notice that, when applying AIC or BIC, $U_1, \dots, U_n$ will not be directly %counted as free parameters because the  information criterion approach requires a distribution modeling on $U's$ in order to get the log-likelihood, such that the free parameters would be of the %modeling distribution. The third and forth terms reflect the model complexity for $\hat A$ and $\hat B$, which are proportional to the noise intensity. Without noise, these two terms would reach %their minimum at $p_0(p-p_0)$ and $q_0(q-q_0)$.

The rank selection method chooses the rank pair $(\hat p_0, \hat q_0)$ to minimize over $(p_0,q_0)$ the criterion
%The practical implementation to calculate $ {\rm df}_{(\tilde p,\tilde q)}$ would remove the  expectation of (\ref{mpca_df}) 
%to get its estimation as $\hat{\rm df}_{(\tilde p,\tilde q)}$.
%In summary,  the selection criterion is to choose the rank pair {$(\hat p,\hat q)$ to minimize} %${\rm IC}(\tilde p,\tilde q,\sigma^2)$ where
%%%%%%%%%%%%%%%%%%%%%%%%%%%%%%%%
%%         Eq 1.8            %%
%%%%%%%%%%%%%%%%%%%%%%%%%%%%%%%%
\begin{align}
{\rm SURE}({p_0, q_0}; \sigma^2) = n^{-1}\sum^n_{k=1}\|\Xb_k-\hat A \hat U_k \hat B^\top\|_F^2 + 2n^{-1}\sigma^2{\rm df}_{(p_0, q_0)}-pq\sigma^2,
\label{SURE}
\end{align}
in which $\sigma^2$ is assumed known or replaced by its estimate $\hat \sigma^2$ descirbed in Section 3.3 and Algorithms 1 and 2 of Tu et al. \cite{Iping2019}. A basic insight underlying high-dimensional covariance matrix estimation is that the commonly used average of tail eigenvalues tends to under-estimate $\sigma^2$ because the empirical distribution (based on a sample of size $n$) of the eigenvalues of $I_p$ converges weakly to the Marchenko-Pastur distribution as $n \rightarrow \infty$ and $p/n$ converges to a positive constant.  For a large random martix, Ulfarsson and Solo \cite{ulfarsson2008dimension} use an upper bound on the number of eigenvalues for its bulk and Tu et al. \cite{Iping2019} modify this idea in their Algorithm 2 for the PCA model, which they then apply to the estimation of $\sigma^2$ in the MPCA model.

\subsection{Rank selection for PCA}
%{\color{blue}We next consider rank selection for the vector $\nu$ of uncorrelated principal components. Chang et al.  \cite[pp. 2442-2443]{chang2014asymptotic} give a brief review of the extensive literature on information criteria, such as AIC and BIC, for variable selection in regression and time series models and point out that although geostatistical regression can be regarded as a linear mixed model, its asymptotic behavior is surprisingly subtle than usual.}
Noting that the commonly used information criterion AIC or BIC for variable selection in regression and time series models can be viewed as an estimator of the Kullback-Leibler 
divergence between the true model and a fitted model under the assumption
that estimation is carried out by maximum likelihood for a parametric family
that includes the true model, Konishi and Kitagawa \cite{konishi1996generalised} introduced a “generalized information criterion” (GIC) to relax the assumption in various
ways. Recently Hung et al. \cite{hung2020generalized} developed GIC for high-dimensional PCA rank selection, which we summarize below and will use in Section 2. A {\it generalized spiked covariance model} for an $m$-dimensional random vector has the spectral decomposition  $\Sigma=\Gamma\Delta\Gamma^\top$ for its covariance matrix $\Sigma$, where $\Delta=diag(\delta_1,\dots,\delta_m)$ with $\delta_1>\cdots>\delta_r \gg \delta_{r+1}>\cdots>\delta_m$. We call $r$ the "generalized rank". The sample covariance matrix $S_n$ has the spectral decomposition $S_n=\Sigma_{j=1}^m \hat \delta_j \hat \gamma_j \hat \gamma_j^\top$. Bai et al. \cite{bai2018consistency} consider the special case of with $\delta_{r+1}=\dots=\delta_m$, called the "simple spiked covariance model" and denoted by $\Sigma_r$. Under the assumption of i.i.d. Gaussian $y_i$, Bai et al. \cite{bai2018consistency} prove consistency of AIC and BIC by using random matrix theory. % remove i.i.d. and delta

Following Konishi and Kitagawa's framework of model selection \cite{konishi1996generalised}, Hung et al. \cite{hung2020generalized} develop GIC for PCA rank selection in possibly misspecified generalized spiked covariance models with generalized rank $r$, for which they use $b_r^{\rm GIC}$ to denote the asymptotic bias correction, a major ingredient of GIC. Their Theorem~2 shows that under the distributional working model assumption of i.i.d. normal $y_i$ with covariance matrix $\Sigma_r$, $b_r^{\rm GIC}$ can be expressed as 
%%%%%%%%%%%%%%%%%%%%%%%%%%%%%%%%
%%         Eq 1.9            %%
%%%%%%%%%%%%%%%%%%%%%%%%%%%%%%%%
\begin{align}\label{GICpen}
b_r^{\rm GIC} &=&
\begin{pmatrix}r\\2
\end{pmatrix} +\sum_{j=1}^{r} \sum_{\ell=r+1}^{ m} \frac{ \delta_\ell(\delta_j- \delta_r )}{\delta_r ( \delta_j-\delta_\ell)} + r + \frac{( m-r)^{-1}\sum^{m}_{j=r+1}\delta^2_j}{\left\{
(m- r)^{-1}\sum^{ m}_{j=r+1}\delta_j
\right\}^2},
\end{align}
and the GIC-based rank selection criterion is 
%%%%%%%%%%%%%%%%%%%%%%%%%%%%%%%%
%%         Eq 1.10            %%
%%%%%%%%%%%%%%%%%%%%%%%%%%%%%%%%
\begin{align}\label{GICpen2}
\hat r_{\rm GIC}=\argmin_{r\le m} \left( \log|\hat\Sigma_r|+{\log n \over n}\hat b_r^{\rm GIC}\right) \text{, where $\hat b_r^{\rm GIC}$ replaces $\delta_i$ by $\hat \delta_i$ in \eqref{GICpen}.\text{\footnotemark}
} 
\end{align}
\footnotetext{The second summand on the right-hand side of (\ref{GICpen}) is actually a slight modification of that in 
\cite{hung2020generalized} to achieve improved performance.
}
%\footnote{
%Note that  \eqref{GICpen} is a modification of the $b_r^{\rm GIC}$  in \cite{hung2020generalized} where we replace the average of the %tail eigenvlaues
%starting from the $(r+1)$-th one by the $r$-th eigenvalue $\delta_r$ 
%in the second summand of \eqref{GICpen} because of better numerical performance.
%}
Note that the last summand in \eqref{GICpen} is $\geq 1$, with equality if and only if  $\delta_{r+1}=\dots=\delta_m$, which is the simple spiked covariance model considered by Bai et al. \cite{bai2018consistency},\footnote{Bai et al. \cite{bai2018consistency} assume that the smallest $m-r$ eigenvalues of the true covariance matrix to be all equal in their proof of rank selection consistency of AIC or BIC.} and that GIC assumes $y_i$ to be a random sample generated from an actual distribution with density function $f$ and uses the Kullback-Leibler (KL) divergence $I(f,g)$ of a "working model" $g$ that is chosen to have the smallest KL divergence from a family of densities; see \cite[pp. 876-879]{konishi1996generalised}.
 %In addition, the last summand %in \eqref{GICpen} is $\geq 1$, %with equality if and only if %$\delta_{r+1}=\dots=\delta_m$, %which is the simple spiked %covariance model considered by %Bai et al. %\cite{bai2018consistency}, %whereas GIC assumes that the %$y_i$ are a random sample %enerated from an actual %distribution with density %function $f$ and uses the %Kullback-Leibler (KL) %divergence $I(f,g)$ of a %"working model" $g$ that is %chosen to have the smallest KL %divergence from a family of %&densities; see \cite[pp. %&876-879]{konishi1996generalise%d}.
 Since $f$ is unknown, Konishi and Kitagawa's idea is to estimate $I(f,g)$ via an "empirical influence function". Hung et al. \cite{hung2020generalized} basically implement this approach in the context of PCA rank selection, for which random matrix theory and the Marchenko-Pastur distribution provide key tools in the setting of high-dimensional covariance matrices. However, even with these powerful tools, the estimate of $I(f,g)$ involves either higher moments of $y_i$ "which can be unstable in practice" or the Stieltjes transform of the limiting Marchenko-Pastur distribution that is difficult to invert to produce explicit formulas for the bias-corrected estimate of $I(f,g)$. Hung et al. \cite{hung2020generalized} preface their Theorem 2 with the comment that "a neat expression of $b_r^{\rm GIC}$ that avoids calculating high-order moments (of the $y_i$) can be derived under the working assumption of (their) Gaussianity" and follow up with Remark 2 and Sections 3.2 and 3.3 to show that this Gaussian assumption "is merely used to get an explicit neat expression for $b_r^{\rm GIC}$" and "is not critical in applying" the rank selection criterion $\hat r_{\rm GIC}$, which is shown in their Theorems 7,8 and 10 to be consistent under conditions that do not require Gaussianity.

\section{Two-stage dimensional reduction (2SDR) for high-dimensional noisy images}
We have reviewed in Sections 1.3--1.5 previous works on PCA and MPCA models, in particular the use of the "Kronecker envelope" ${\rm span}(B \otimes A)$ in~\eqref{eq:PCA} as an attempt to connect both models. This attempt, however, is incomplete because it does not provide an explicit algorithm to compute the "full rank $p_0q_0 \times r$ matrix $G$" that is shown to "exist". In this section we define a new model , called {\it hybrid PCA} and denoted by ${\rm H_M PCA}$, in which the subscript M stands for MPCA and H stands for "hybrid" of MPCA and PCA.  Specifically, ${\rm H_M PCA}$ assumes the MPCA model~\eqref{eq:MPCA} with reduced rank $(p_0,q_0)$ via the $p_0 \times q_0$ random matrix $U$ and then assumes a rank-$r$ model, with $r \le p_0q_0$, for $\vec(U)$ to which 
a zero-mean random error $\epsilon$ with ${\rm Cov}(\epsilon)=cI_{p_0,q_0}$ is added, as in~\eqref{eq:PCA_vec}. This leads to dimension reduction of $\vec(X-M-\cE)=\vec(A_{p_0}UB_{q_0}^{\top})$ from $p_0q_0$ to $r$. Since $U=A_{p_0}^{\top}(X-M)B_{p_0}$ in view of~\eqref{eq:MPCA_vec}, $\vec(A_{p_0}UB_{q_0}^{\top})=P_{B_{q_0} \otimes A_{p_0}}\vec(X-M-\cE)$ is the projection of $X-M-\cE$ into ${\rm span}(B_{q_0} \otimes A_{p_0})$, which has dimension $r$ after this further rank reduction. The actual ranks, which we denote by $(p_0^*, q_0^*)$ and $r^*$, are unknown as are the other parameters of the ${\rm H_M PCA}$ model, and 2SDR uses a sample of size $n$ to fit the model and estimate the ranks.

The first stage of 2SDR uses \eqref{eq:MPCA} to model a noisy image $X$ as a matrix. Ye's estimates $\hat A$ and $\hat B$ that we have described in the second paragraph of Section 1.3 depend on the given value of $(p_0, q_0)$, which is specified in the criterion ${\rm SURE}(p_0, q_0; \sigma^2)$. Direct use of the criterion to search for $(p_0, q_0)$ would therefore involve computation-intensive loops. To circumvent this difficulty, we make use of the analysis of the optimization problem associated with Ye's estimates at the population level by Hung et al. \cite{BioHung2012} and Tu et al. \cite[pp. 27-28]{Iping2019}, who have shown that if the dimensionality is over-specified (respectively, under-specified) for span$(A)$ (or span$(B)$), then it contains (respectively, is a proper subspace of) the true subspace. We therefore choose a rank pair $(p_u, q_u)$  such that $p_u \geq p_0^*$ and $q_u \geq q_0^*$, where $(p_0^*, q_0^*)$ is the true value of $(p_0, q_0)$, and then solve for $(\hat A_u,\hat B_u)$ such that $\hat A_u$ (respectively, $\hat B_u$) consists of the leading $p_u$ eigenvectors of $\sum_{k=1}^n (X_k -\overline X) P_{\hat B_u} (X_k-\overline X)^\top$ (respectively, $q_u$ eigenvectors of $ \sum_{k=1}^n (X_k -\overline X) P_{\hat A_u} (X_k-\overline X)^\top$); see Ye's estimate in the second paragraph of Section 1.3. This is tantamout to replacing $(p_0^*, q_0^*)$ by the larger surrogates $(\hat p_u, \hat q_u)$ in ${\rm df}_{(p_0^*,q_0^*)}$ defined by \eqref{mpca_df}. For given $(p_0,q_0)$ with $p_0 \le p_u$ and $q_0 \le q_u$, let $\hat A_{p_0}$ (respectively, $\hat B_{q_0}$) be the submatrix consisting of the first $p_0$ (respectively, $q_0$) column vectors of $\hat A_u$ (respectively, $\hat B_u$).   The rank selection criterion \eqref{SURE} into the SURE criterion is 
\begin{align}
{\rm S}_u^{(n)}(p_0,q_0 ;\sigma^2) = n^{-1}\sum^n_{k=1}\|\Xb_k-\widehat A_{p_0} \widehat U_k \widehat B_{q_0}^\top \|_F^2 + 2n^{-1}\sigma^2{\rm df}^{(n)}_{ p_0, q_0}-pq\sigma^2, \ \mbox{where}
\label{ICC}
\end{align}
%%%%%%%%%%%%%%%%%%%%%%%%%%%%%%%%
%%         Eq 2.2            %%
%%%%%%%%%%%%%%%%%%%%%%%%%%%%%%%%
\begin{align}\label{df22}
{\rm df}^{(n)}_{p_0,q_0} = pq+(n-1)p_0q_0 +
\sum_{i=1}^{p_0} \sum_{\ell=p_0+1}^{p} \frac{\widehat \lambda_i +
\widehat \lambda_\ell}{\widehat \lambda_i - \widehat \lambda_\ell}
+
\sum_{j=1}^{q_0} \sum_{\ell=q_0+1}^{q} \frac{\widehat \xi_j +\widehat \xi_\ell}{\widehat \xi_j - \widehat \xi_\ell},
\end{align}
in which $\widehat U_k= \widehat A_{p_0}^\top(X_k-\overline X) \widehat B_{q_0}$ and %the $\lambda_i$ (respectively, $\xi_i$) are the eigenvalues, in decreasing order of their magnitudes, of  $n^{-1}\sum_{k=1}^n (X_k -\overline X) P_{\hat B} %(X_k-\overline X)^\top$ (respectively, of  $n^{-1}\sum_{k=1}^n (X_k -\overline X) P_{\hat A} %(X_k-\overline X)^\top$), 
%$\tau_n$ represents the "penalty" that depends on the %sample size $n$, with $\tau_n \rightarrow \infty$ and %$n^{-1/2}\tau_n \rightarrow 0$ as $n \rightarrow \infty$ %in contrast to $\tau_n=1$ in \eqref{mpca_df}.} Define
%%%%%%%%%%%%%%%%%%%%%%%%%%%%%%%%
%%         Eq 2.3            %%
%%%%%%%%%%%%%%%%%%%%%%%%%%%%%%%%
\begin{align}\label{first_stage}
(\hat p_0, \hat q_0)={\rm argmin}_{p_0 \le p_u, q_0 \le q_u} 
{\rm S}_u^{(n)}(p_0,q_0;\sigma^2). 
\end{align}
In Section 2.1, we prove the consistency of $(\hat p_0, \hat q_0)$ as an estimate of $(p_0^*, q_0^*)$.

%PCA model \eqref{eq:PCA_vec} to $\VEC(\Ub) \in %\mathbb{R}^{p_0^*q_0^*}$:
%%%%%%%%%%%%%%%%%%%%%%%%%%%%%%%%%
%%%         Eq 2.4            %%
%%%%%%%%%%%%%%%%%%%%%%%%%%%%%%%%%
%\begin{equation}
%\begin{aligned}\label{eq:hatV}
%\VEC(\Ub)&=\Gb\nu+\epsilon, 
%\end{aligned}
%\end{equation}
%in which $\nu \in \mathbb{R}^r$ is a zero-mean random vector with %uncorrelated components and is independent of $\epsilon$ that has %mean 0 and ${\rm Cov}(\epsilon)=cI_{p_0^*q_0^*}$, and $G$ is a %nonrandom $p_0^*q_0^* \times r$ matrix with orthogonal column %vectors. Hence $\VEC(U)$ has a generalized spiked covariance matrix %of the form
%where $\kappa_1\geq \dots \geq\kappa_{r} > c > 0$ and the $g_j$ are %orthonormal column vectors. %Since $\VEC(X) = \VEC(M)+{(B \otimes %A)}\VEC(U)+\VEC(\cE)$ 
%In view of \eqref{eq:MPCA} and  ${\rm Cov}(\VEC(\cE)) = \sigma^2 %I_{pq}$, it follows that the column vectors
%of $(B \otimes A)G$ compose the first $r$ eigenvectors of ${\rm %Cov}(\vec(X))$ and the whole spectrum of eigenvalues of ${\rm %Cov}(\vec(X))$ are

After rank reduction via consistent estimation of $(p_0^*, q_0^*)$ by $(\hat p_0, \hat q_0)$, the second stage of 2SDR achieves further rank reduction by applying the GIC-based rank selection criterion~\eqref{GICpen2} to $S_{\vec(\hat U)}=n^{-1}\sum_{i=1}^n \vec(\hat U_i)\vec( \hat U_i)^{\top}$, where $\hat U_i=\hat A_{\hat p_0}^{\top} (X_i-\overline X)\hat B_{\hat q_0}$. Consider the corresponding matrix $\Sigma_{\vec(U)}$ at the population level and its ordered eigenvalues $\kappa_1\geq \kappa_{2} \geq \dots$ and the orthonormal eigenvectors $g_j$ associated with $ \kappa_j$. Suppose $\vec(U-\epsilon)$ belongs to an $r$-dimensional subspace. Since $\epsilon$ is independent of $\vec(U-\epsilon)$ and has mean 0 and covariance matrix $cI_{p_0^*q_0^*}$, it then follows that 

%PCA model \eqref{eq:PCA_vec} to $\VEC(\Ub) \in %\mathbb{R}^{p_0^*q_0^*}$:
%%%%%%%%%%%%%%%%%%%%%%%%%%%%%%%%%
%%%         Eq 2.4            %%
%%%%%%%%%%%%%%%%%%%%%%%%%%%%%%%%%
%\begin{equation}
%\begin{aligned}\label{eq:hatV}
%\VEC(\Ub)&=\Gb\nu+\epsilon, 
%\end{aligned}
%\end{equation}
%in which $\nu \in \mathbb{R}^r$ is a zero-mean random vector with %uncorrelated components and is independent of $\epsilon$ that has %mean 0 and ${\rm Cov}(\epsilon)=cI_{p_0^*q_0^*}$, and $G$ is a %nonrandom $p_0^*q_0^* \times r$ matrix with orthogonal column %vectors. Hence $\VEC(U)$ has a generalized spiked covariance matrix %of the form
%%%%%%%%%%%%%%%%%%%%%%%%%%%%%%%%
%%         Eq 2.5            %%
%%%%%%%%%%%%%%%%%%%%%%%%%%%%%%%%
\begin{equation}\label{2SDR-PCA}
\Sigma_{\vec(U)}=  \sum^{r}_{j=1}\kappa_j\gb_j\gb_j^\top +c \sum^{p_0^*q_0^*}_{j=r+1}\gb_j\gb_j^\top,
\end{equation}
 Since $X_i=M+A_{p_0*}U_iB_{q_0*}^{\top}+\cE_i$ with ${\rm Cov}(\cE_i)=\sigma^2I_{pq}$ in view of~\eqref{eq:MPCA}, it then follows that the covariance matrix  of $\vec(X_i), 1 \le i \le n$, has eigenvalues
%where $\kappa_1\geq \dots \geq\kappa_{r} > c > 0$ and the $g_j$ are %orthonormal column vectors. %Since $\VEC(X) = \VEC(M)+{(B \otimes %A)}\VEC(U)+\VEC(\cE)$ 
%In view of \eqref{eq:MPCA} and  ${\rm Cov}(\VEC(\cE)) = \sigma^2 %I_{pq}$, it follows that the column vectors
%of $(B \otimes A)G$ compose the first $r$ eigenvectors of ${\rm %Cov}(\vec(X))$ and the whole spectrum of eigenvalues of ${\rm %Cov}(\vec(X))$ are

%%%%%%%%%%%%%%%%%%%%%%%%%%%%%%%%
%%         Eq 2.6            %%
%%%%%%%%%%%%%%%%%%%%%%%%%%%%%%%%
\begin{equation}\label{2SDR-eigev}
\kappa_1+\sigma^2,\dots,  \kappa_{r}+\sigma^2,  
\underbrace{c+\sigma^2, \cdots c+\sigma^2}_{\text{$p_0^*q_0^*-r$ times}}, \underbrace{\sigma^2,\cdots, \sigma^2}_{\text{$pq-p_0^*q_0^*$ times}}.
\end{equation}
%which can be used to compute $\log|\hat \Sigma_r|$ after replacing $p_0^*q_0^*$ by $\hat p_0\hat q_0$.
Although $p_0^*,q_0^*,r,\sigma^2,c$ and other parameters such as $\kappa_1, \cdots, \kappa_r$ are actually unknown in \eqref{2SDR-PCA} and \eqref{2SDR-eigev}, 2SDR uses a sample of size $n$ to fit the model and thereby obtain the estimate of $\kappa_1, \cdots, \kappa_r$, which can be put into $\log|\hat \Sigma_r|$ after replacing $p_0^*q_0^*$ by $\hat p_0\hat q_0$; details are given in  Section 2.1 and illustrated in Section 2.2.

%%%%%%%%%%%%%%%%
%% Figure 1   %%
%%%%%%%%%%%%%%%%
\begin{figure}[]
\centering
\includegraphics[width=0.85\textwidth]{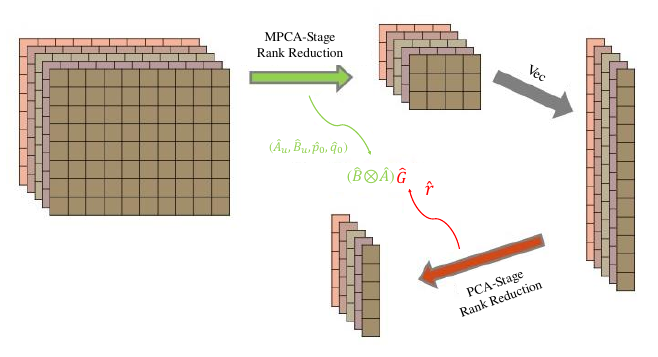} 
\caption{Schematic of 2SDR procedure.}
\label{fig:2SDR}
\end{figure}

%\vspace*{\floatsep}

%%%%%%%%%%%%%%%%%%%%%%%%%%%%%%%%
%%         Section 2.1        %%
%%%%%%%%%%%%%%%%%%%%%%%%%%%%%%%%
\subsection{Implementation and theory of 2SDR}
The first paragraph of this section has already given details to implement the MPCA stage of 2SDR that yields $\hat A_u, \hat B_u, \hat A_{p_0},  \hat B_{q_0}, p_u, q_u, \hat p_0$ and $\hat q_0$, hence this subsection will provide details for the implementation of the basic idea underlying the second (PCA) stage of 2SDR described in the preceding paragraph and develop an asymptotic theory of 2SDR, particularly the consistency of the rank estimates in the ${\rm H_mPCA}$ model. Figure \ref{fig:2SDR} is a schematic diagram of the 2SDR procedure, which consists of 4 steps, the first two of which constitute the MPCA stage while the last two form the second (PCA) stage of 2SDR, based on a sample of $p \times q$ matrices $X_i \ (i=1, \cdots, n)$ representing high-dimensional noisy images.
%To select $(\hat p_0, \hat q_0, \hat r)$, we apply the following steps:
\begin{itemize}[leftmargin=*]
\item \textbf{Step 1.}  Fit the MPCA model \eqref{eq:MPCA} to the sample of size $n$ to obtain the eigenvalues $\hat\lambda_1,\cdots,\hat\lambda_{p}$, $\hat\xi_1,\cdots,\hat\xi_{q}$ in \eqref{df22} and the matrices $\hat A_u$ and $\hat B_u$.
\item \textbf{Step 2.}
Estimate $\sigma^2$ by the method of Tu et al. \cite{Iping2019}\footnote{We use 7/8 from the tail to estimate the variance of noise for the SURE method.} described in the last paragraph of Section 1.4 and use the SURE rank selection criterion \eqref{first_stage} to choose the reduced rank $(\hat p_0, \hat q_0)$.
\item \textbf{Step 3.} Perform PCA on  $S_{\vec(\hat U)}=n^{-1}\sum_{i=1}^n\vec(\hat U_i)\vec(\hat U_i)^\top$ to obtain its ordered eigenvalues $\hat\kappa_1 \ge \cdots \hat \kappa_{\hat p_0 \hat q_0}$.
 %\cite{hung2020generalized}.
\item \textbf{Step 4.} For $r \le \hat p_0 \hat q_0$, define ~\eqref{2SDR-PCA} with $\kappa_j$ replaced by $\hat \kappa_j$, and compute $\log|\hat \Sigma_r|$ and  $\hat b_r^{{\rm GIC}}$. Estimate the actual rank $r^*$ by $\hat r_{{\rm GIC}}$ defined in \eqref{GICpen2}.   %\cite{hung2020generalized}.
\end{itemize}

Hung et al. \cite[Corollary 1]{BioHung2012} have shown that $\vec(P_{\hat B \otimes \hat A})$ is a $\sqrt{n}$-consistent estimate of $\vec(P_{B \otimes A})$ under certain regularity conditions, which we use to prove the following theorem on the consistency of $(\hat p_0, \hat q_0)$. The theorem needs the condition $c> \sigma^2 $ because $2n^{-1}\sigma^2{\rm df}^{(n)}_{p_0,q_0}$ in (\ref{ICC}) can dominate the value of the criterion ${\rm S}_u^{(n)}(p_0,q_0; \sigma^2)$, especially when $p_0 < p_0^*$ and $q_0 <q_0^* $. The condition $c > \sigma^2$, or equivalently $c+\sigma^2 >2\sigma^2$, ensures sufficiently large  ${\rm S}_u^{(n)}(p_0,q_0; \sigma^2)$ to avoid erroneous rank selection in this case. The asymptotic analysis, as $n \rightarrow \infty$, in the following proof also shows how the condition $c > \sigma^2$ is used.

\begin{theorem*}\label{thm1} Assume that the MPCA model 
(\ref{eq:MPCA}) holds and $ c > \sigma^2$. Suppose $p_0^* \leq p_{\pre}$ and $q_0^* \leq q_{\pre}$. Then $\mathrm{P}(\hat p_0= p_0^*\mbox{~~and~~}\hat q_0 = q_0^*)\rightarrow 1\mbox{~~as~~}n\rightarrow \infty.
$
\end{theorem*}
%%%%%%%%%%%%%%%%%%%%%%%%%%%%%%%%
%%         Proof        %% 
%%%%%%%%%%%%%%%%%%%%%%%%%%%%%%%%
\begin{proof}
The basic idea underlying the proof is the 
decomposition for the rank pair $(p_0,q_0)$ with 
$p_0 \leq p_{\pre}$ and $q_0 \leq q_{\pre}$:
\begin{align}
{\rm S}_u^{(n)}(p_0,q_0; \sigma^2) - {\rm S}_u^{(n)}(p^*_0,q^*_0; \sigma^2)=A_{1n}+A_{2n}+A_{3n}
\label{ic1_0}
\end{align}
consisting of three summands which can be analyzed separately by using different arguments 
to show that (\ref{ic1_0}) is 
{ (a) at least of the order $n^{-1/2}$ in probability, (b) equal to $1+o_P(1)$ if $p_0< p_0^*$ or $q_0< q_0^*$, and (c) of the $O_P(n^{-1/2}\tau_n)$ order if $p_0 \geq p_0^*$ and $q_0 \geq q_0^*$}, ensuring that (\ref{ic1_0}) is
sufficiently above zero if $p_0 \neq p^*_0$ or $q_0 \neq q^*_0$.
In view of (\ref{ICC}), the left-hand side of (\ref{ic1_0})
is equal to 
\begin{align}
\frac{1}{n}\sum^n_{i=1}
\left\{
\|\Xb_i -\widehat \Ab_{p_0}
\widehat \Ub_i \widehat B_{q_0}
\|_F^2  - \|\Xb_i -\widehat \Ab_{p_0^*}
\widehat \Ub_i \widehat \Bb_{q_0^*}
\|_F^2 \right\} +A_{3n},~A_{3n}= \frac{2\sigma^2}{n}
({\rm df}^{(n)}_{p_0q_0} - {\rm df}^{(n)}_{p_0^*q^*_0}). \label{decomp}
\end{align}
We  next show that the first summand in (\ref{decomp}) is equal to $A_{1n}+A_{2n}$, where 
\begin{eqnarray*}
A_{1n} &=& \frac{1}{n}\sum^n_{i=1}\|(\Pb_{\widehat\Bb_{q_0^*}}\otimes \Pb_{\widehat\Ab_{p_0^*}} -\Pb_{\widehat\Bb_{q_0}}\otimes \Pb_{\widehat\Ab_{p_0}} )\vect(\Xb_i)\|^2_F\notag\\
A_{2n} &=& 
 \frac{2}{n}\sum^n_{i=1}\langle (\Ib_{pq}-\Pb_{\widehat\Bb_{q_0^*}}\otimes \Pb_{\widehat\Ab_{p_0^*}} )\vect(\Xb_i),~
(\Pb_{\widehat\Bb_{q_0^*}}\otimes \Pb_{\widehat\Ab_{p_0^*}} -\Pb_{\widehat\Bb_{q_0}}\otimes \Pb_{\widehat\Ab_{p_0}} )\vect(\Xb_i)
\rangle,
\end{eqnarray*}
by writing $n^{-1} 
\displaystyle\sum^n_{i=1} \|\Xb_i -\widehat \Ab_{p_0}\widehat \Ub_i \widehat \Bb_{q_0}^\top\|_F^2$ as 
\begin{align*}
&\frac{1}{n}\sum^n_{i=1}\|\Xb_i - \Pb_{\widehat\Ab_{p_0^*}}\Xb_i
\Pb_{\widehat\Bb_{q_0^*}}+(\Pb_{\widehat\Ab_{p_0^*}} \Xb_i
\Pb_{\widehat\Bb_{q_0^*}}-
\Pb_{\widehat\Ab_{p_0}}\Xb_i
\Pb_{\widehat\Bb_{q_0}})\|^2_F\notag\\
&=\frac{1}{n}\sum^n_{i=1}\|(\Ib_{pq}-\Pb_{\widehat\Bb_{q_0^*}}\otimes \Pb_{\widehat\Ab_{p_0^*}} )\vect(\Xb_i)+(\Pb_{\widehat\Bb_{q_0^*}}\otimes \Pb_{\widehat\Ab_{p_0^*}} -\Pb_{\widehat\Bb_{q_0}}\otimes \Pb_{\widehat\Ab_{p_0}} )\vect(\Xb_i)\|^2_F\notag\\
&=\frac{1}{n}\sum^n_{i=1} \|\Xb_i -\widehat \Ab_{p_0^*}\widehat \Ub_i \widehat \Bb_{q_0^*}^\top\|^2_F+A_{1n}+
A_{2n}.
\end{align*}

Let $\Sigmab$ be the covariance matrix of $\vect(\Xb)$, $\displaystyle\widehat \Sigma=  n^{-1}\sum^n_{i=1} \vect(\Xb_i)\vect(\Xb_i)^\top$, and note that 
$\widehat\Sigmab=\Sigmab +O_P(n^{-1/2})$. By  Corollary 1 of \cite{BioHung2012}, 
%\color{red}
 \begin{align}\label{hung_eq2}
 \widehat \Ab_{p_0^*} &= \Ab_{p_0^*} +O_P(n^{-1/2}),~\widehat \Bb_{q_0^*}= \Bb_{q_0^*} + O_P(n^{-1/2}),\\~
\widehat \Ab_{p_0} &= \Ab_{p_0} +O_P(n^{-1/2}),~ \widehat \Bb_{q_0} = \Bb_{q_0} +O_P(n^{-1/2}), \mbox{ for } 1\leq p_0\leq p_u \mbox{ and } 1\leq q_0\leq q_u.   \notag
\end{align}
 Note also that 
$A_{2n} =
2{\rm tr}([\Pb_{\widehat\Bb_{(q_0^*\wedge q_0)}}
\otimes \Pb_{\widehat\Ab_{(p_0^*\wedge p_0)}} -
\Pb_{\widehat\Bb_{q_0}}
\otimes \Pb_{\widehat\Ab_{p_0}}]\widehat\Sigmab)$. From this and (\ref{hung_eq2}), it follows that  
\begin{align}
A_{2n} =2\sigma^2 \{(p_0^*\wedge p_0)(q_0^*\wedge q_0)-p_0q_0 \}+O_P(n^{-1/2}). \label{A2n}
\end{align}
Similar calculations show that 
$ A_{1n}=   {\rm tr}((\Pb_{\widehat\Bb_{q_0^*}}\otimes \Pb_{\widehat\Ab_{p_0^*}})\widehat\Sigmab)
+ {\rm tr}((\Pb_{\widehat\Bb_{q_0}}\otimes \Pb_{\widehat\Ab_{p_0}})\widehat\Sigmab) \notag\\
- 2{\rm tr}((\Pb_{\widehat\Bb_{q_0 \wedge q_0^*}}\otimes \Pb_{\widehat\Ab_{(p_0 \wedge p_0^*)}})\widehat\Sigmab)
+O_P(n^{-1/2})$, to which ( \ref{hung_eq2}) can be 
applied to obtain 
\begin{eqnarray}\label{new1}
A_{1n}
&=&{\rm tr}([\Ib_{p_0^*q_0^*} - (\Bb^\top\Pb_{\Bb_{q_0}}\Bb)\otimes(\Ab^\top \Pb_{\Ab_{p_0}}\Ab)]\Sigmab_{\vec(\Ub)})\\
&&+\sigma^2 (p_0^*q_0^* + p_0q_0-2(p_0^*\wedge p_0)(q_0^*\wedge q_0))
+O_P(n^{-1/2}). \notag
\end{eqnarray}

By the assumption $c > \sigma^2$,  $\Sigmab_{\vec(U)} - \sigma^2 I_{p^*_0q^*_0}$ is positive definite. By
von Neumann's trace inequality (cf. \cite{Wei2015} ), 
\begin{align}\label{new2}
 \mbox{1st summand in } ~\eqref{new1}
%&&{\rm tr}\left[\{\Ib_{p_0^*q_0^*} - (\Bb^\top\Pb_{\widehat\Bb_{q_0}}\Bb)\otimes(\Ab^\top \Pb_{\widehat\Ab_{p_0}}\Ab)\}\Sigmab_{\Ub}\right] \\
&\geq 
{\rm tr}(\{\Ib_{p_0^*q_0^*} - (\Bb^\top\Pb_{\widehat\Bb_{q_0}}\Bb)\otimes(\Ab^\top \Pb_{\widehat\Ab_{p_0}}\Ab)\}\sigma^2\Ib_{p^*_0q_0^*})\\
&=
\sigma^2 (p_0^*q_0^* -(p_0^*\wedge p_0)(q_0^*\wedge q_0)). \notag 
\end{align}
In the first inequality of (\ref{new2}), 
``='' holds only when $p_0\geq p_0^*$ and $q_0 \geq q_0^*$. 
Note that $\Ib_{p_0^*q_0^*} - (\Bb^\top\Pb_{\widehat\Bb_{q_0}}\Bb)\otimes(\Ab^\top \Pb_{\widehat\Ab_{p_0}}\Ab)$ is   
symmetric and nonnegative definite. 
By (\ref{new1}) and (\ref{new2}), 
\begin{align}
A_{1n}=
\sigma^2 (2p_0^*q_0^* +p_0q_0-3(p_0^*\wedge p_0)(q_0^*\wedge q_0))+O_P(n^{-1/2}).\label{A1n} 
\end{align}
Moreover, $A_{3n}$ in (\ref{decomp}) is equal to
\begin{align*}
&\frac{2(n-1)\sigma^2}{n}(p_0q_0 - p_0^*q_0^*) 
+
\frac{2\sigma^2}{n}
\left[
\sum_{j=1}^{p_0} \sum_{\ell=p_0+1}^{p} \frac{\widehat \lambda_j +
\widehat \lambda_\ell}{\widehat \lambda_j - \widehat \lambda_\ell}
+
\sum_{j=1}^{q_0} \sum_{\ell=q_0+1}^{q} \frac{\widehat \xi_j +\widehat \xi_\ell}{\widehat \xi_j - \widehat \xi_\ell}
\right]\notag\\
&- \frac{2\sigma^2}{n}
\left[
\sum_{j=1}^{p_0^*} \sum_{\ell=p_0^*+1}^{p} \frac{\widehat \lambda_j +
\widehat \lambda_\ell}{\widehat \lambda_j - \widehat \lambda_\ell}
+
\sum_{j=1}^{q_0^*} \sum_{\ell=q_0^*+1}^{q} \frac{\widehat \xi_j +\widehat \xi_\ell}{\widehat \xi_j - \widehat \xi_\ell}
\right],
\end{align*}
which can be combined with (\ref{A2n}), (\ref{A1n}) and (\ref{ic1_0}) to yield 
\begin{align}\label{r2}
&{\rm S}_u^{(n)}(p_0,q_0; \sigma^2) - {\rm S}_u^{(n)}(p^*_0,q^*_0; \sigma^2) \geq \sigma^2(p_0q_0 - (p_0^*\wedge p_0)(q_0^*\wedge q_0)) \\
&+
\frac{2\sigma^2}{n}
\left[
\sum_{j=1}^{p_0} \sum_{\ell=p_0+1}^{p} \frac{\widehat \lambda_j +
\widehat \lambda_\ell}{\widehat \lambda_j - \widehat \lambda_\ell}
+
\sum_{j=1}^{q_0} \sum_{\ell=q_0+1}^{q} \frac{\widehat \xi_j +\widehat \xi_\ell}{\widehat \xi_j - \widehat \xi_\ell}
\right]\notag\\
&-  \frac{2\sigma^2}{n}
\left[
\sum_{j=1}^{p_0^*} \sum_{\ell=p_0^*+1}^{p} \frac{\widehat \lambda_j +
\widehat \lambda_\ell}{\widehat \lambda_j - \widehat \lambda_\ell} +
\sum_{j=1}^{q_0^*} \sum_{\ell=q_0^*+1}^{q} \frac{\widehat \xi_j +\widehat \xi_\ell}{\widehat \xi_j - \widehat \xi_\ell}
\right]+O_P(n^{-1/2}). \notag
\end{align}
{
Note that 
$$\displaystyle\sum_{j=1}^{p_0} \sum_{\ell=p_0+1}^{p} \frac{\widehat \lambda_j +
\widehat \lambda_\ell}{\widehat \lambda_j - \widehat \lambda_\ell}
- \sum_{j=1}^{p_0^*} \sum_{\ell=p_0^*+1}^{p} \frac{\widehat \lambda_j +
\widehat \lambda_\ell}{\widehat \lambda_j - \widehat \lambda_\ell}
 = \sum_{j=p^*_0+1}^{p_0}\sum^{p}_{\ell=j+1} 
\frac{\widehat \lambda_j +
\widehat \lambda_\ell}{\widehat \lambda_j - \widehat \lambda_\ell} 
 \geq 0\mbox{~~if $p_0 >p_0^*$};$$ 
$$\displaystyle\sum_{j=1}^{q_0} \sum_{\ell=q_0+1}^{q} \frac{\widehat \xi_j +\widehat \xi_\ell}{\widehat \xi_j - \widehat \xi_\ell}
- \sum_{j=1}^{q_0^*} \sum_{\ell=q_0+1}^{q} \frac{\widehat \xi_j +\widehat \xi_\ell}{\widehat \xi_j - \widehat \xi_\ell}
 = \sum_{j=q_0^*+1}^{q_0} \sum_{\ell=j+1}^{q} \frac{\widehat \xi_j +\widehat \xi_\ell}{\widehat \xi_j - \widehat \xi_\ell}
 \geq 0\mbox{~~if $q_0 >q_0^*$}.$$  
 On the other hand, if $p_0 < p_0^*$ (respectively, $q_0<q_0^*$), 
 $(\widehat \lambda_j + \widehat \lambda_\ell)/(\widehat \lambda_j - \widehat \lambda_\ell)$ (respectively, $(\widehat \xi_j + \widehat \xi_\ell)/(\widehat \xi_j - \widehat \xi_\ell)$) is positive and bounded away from 0 for  $j \leq p_0$ and $~\ell >p_0$ (respectively,  $j \leq q_0$ and $~\ell >q_0$). Thus, we obtain from  (\ref{r2}) that
 }

%{\color{blue} 
%Since the second line of (\ref{r2}) is positive and 
%$\widehat \lambda_j - \widehat \lambda_\ell$ (respectively, $\widehat \xi_j - %\widehat \xi_\ell$) is away from zero, for $j \leq p_0^*$ and $~\ell >p_0^*$ %(respectively,   $j \leq q_0^*$ and $~\ell >q_0^*$), we obtain from  (\ref{r2}) %that}
\begin{align*}
{\rm S}_u^{(n)}(p_0,q_0; \sigma^2) - {\rm S}_u^{(n)}(p^*_0,q^*_0; \sigma^2)
\geq \sigma^2(p_0q_0 - (p_0^*\wedge p_0)(q_0^*\wedge q_0))+O_P(n^{-1/2}), 
\end{align*}
with equality only when $p_0\geq p_0^*$ and $q_0 \geq q_0^*$, from which it follows that as $n \rightarrow \infty$, 
%$(p_0q_0 - (p_0^*\wedge p_0)(q_0^*\wedge q_0))$ is always non-negative, we have
\begin{align}
{\rm S}_u^{(n)}(p_0,q_0; \sigma^2) - {\rm S}_u^{(n)}(p^*_0,q^*_0; \sigma^2) \left\{
\begin{array}{ll}
=0&  \mbox{~if~} p_0 = p_0^* \mbox{~~and~~} q_0 = q_0^*\\
> o_P(1) & \mbox{~otherwise.} 
\end{array}\right.. 
\label{bb_neq0}
\end{align}
In ~\eqref{bb_neq0}, "${\rm s}_n >o_P(1)$" means that given $\epsilon > 0$ and $\delta > 0$, there exists $n_{\epsilon,\delta}$ such that ${\rm P}(s_n > \epsilon) \geq 1-\delta$ for $n \geq n_{\epsilon, \delta}$, which completes the proof. 
\end{proof}

Since $(p_0^*, q_0^*)$ and $P_{B_{q_0^*}\otimes A_{p_0^*}}$ can be consistently estimated in the first stage of 2SDR, we can apply the consistency of $\hat r_{{\rm GIC}}$ established by Hung et al. \cite{hung2020generalized} for PCA models to obtain the following.

\begin{corollary*}
\label{coro2} Assume that the ${\rm H_MPCA}$ model holds and $c > \sigma^2$. Suppose
that $r^* \leq r_{\pre}$. Then $ \mathrm{P}(\hat r_{\rm GIC}  =  r^*)\rightarrow 1\mbox{~~as~~}n\rightarrow \infty.$ %{\rm (b) }$\mathrm{P}(\hat r_{\rm BIC}  =  r_0)\rightarrow 1\mbox{~~as~~}n\rightarrow \infty$.$
\end{corollary*}
%in (\ref{eq:2SDR}) 
%\begin{proof}
%{ The proof can be found in \cite{hung2020generalized}}. 
%\end{proof}
%\noindent{\red Notice that Corollary requires the condition $c> \sigma^2$ since 
%the second stage, PCA stage, follows the result of the first stage,  MPCA stage. }

%%%%%%%%%%%%%%%%%%%%%%%%%%%%%%%%
%%         Section 2.2        %%
%%%%%%%%%%%%%%%%%%%%%%%%%%%%%%%%
\subsection{Simulation study of performance}\label{s:sim_cryoem} 

\begin{table}[!b]
\caption[Caption for LOF]{MSE, with standard deviation in parentheses, of PCA, MPCA and 2SDR.}%\footnotemark}
\begin{subtable}[h]{\textwidth}
\caption{Data generated from PCA model ~\eqref{eq:setting}.}
\begin{tabular}{c|c|c|c|c|c|c}
\hline
                     & PCA      & MPCA      & 2SDR     & PCA      & MPCA      & 2SDR     \\ \hline
 \backslashbox{$c$\kern-2em}{n}                   & \multicolumn{3}{c|}{1000}         & \multicolumn{3}{c}{100}          \\ \hline
 %\multirow{3}{*}{4}  & \multicolumn{3}{c|}{$Setting~1:$} & \multicolumn{3}{c}{$Setting~3:$} \\ %\cline{1-1} \cline{3-8} 
 \multirow{2}{*}{4}                     & 0.1831     & 1.9851    & 1.8070   & 1.2297     & 2.0951    & 1.9694   \\ %\cline{1-1} \cline{3-8} 
                & (0.0011)     & (0.0172)    & (0.0172)   & (0.0089)     & (0.0562)    & (0.0562)   \\ \hline
 %\multirow{3}{*}{20} & \multicolumn{3}{c|}{$Setting~2:$} & \multicolumn{3}{c}{$Setting~4:$} \\ %\cline{1-1} \cline{3-8} 
 \multirow{2}{*}{20}      & 1.1695     & 3.1956    & 2.2868   & 6.8777     & 3.8893    & 3.1972   \\ %\cline{1-1} \cline{3-8} 
                & (0.0078)     & (0.0199)    & (0.0188)   & (0.0461)     & (0.0585)    & (0.0576)   \\ \hline
\end{tabular}

\end{subtable}
\bigskip
\begin{subtable}[h]{\textwidth}
\caption{Data generated from $\rm H_MPCA$ model ~\eqref{eq:setting2}.}
%\label{tbl:sim_mse_2sdr_tail}
\begin{tabular}{c|c|c|c|c|c|c}
\hline
                 & PCA      & MPCA      & 2SDR     & PCA      & MPCA      & 2SDR     \\ \hline
\backslashbox{$\sigma^2$}{n}                   & \multicolumn{3}{c|}{1000}         & \multicolumn{3}{c}{100}          \\ \hline
%\multirow{3}{*}{1.1}  & \multicolumn{3}{c|}{$Setting~1:$} & \multicolumn{3}{c}{$Setting~3:$} \\ %\cline{1-1} \cline{3-8}
\multirow{2}{*}{1.1}                 & 0.0174         & 0.0578           & 0.0089          & 0.1105            & 0.0705           & 0.0255             \\ %\cline{1-1} \cline{3-8} 
            & (0.0001)         & (0.0003)           & (0.0001)          & (0.0011)            & (0.0010)           & (0.0001)            \\     \hline
%\multirow{3}{*}{5.6} & \multicolumn{3}{c|}{$Setting~2:$} & \multicolumn{3}{c}{$Setting~4:$} \\ %\cline{1-1} \cline{3-8} 
\multirow{2}{*}{5.6}                 & 0.0937         & 0.2947           & 0.0457          & 0.6150            & 0.3662           & 0.1363             \\ %\cline{1-1} \cline{3-8} 
            & (0.0008)         &  (0.0016)          & (0.0005)          & (0.0073)            & (0.0057)           & (0.0031)             \\     \hline
%                                       & \multirow{3}{*}{21} & \multicolumn{3}{c|}{$Setting~3:$} & \multicolumn{3}{c}{$Setting~6:$} \\ \cline{1-1} \cline{3-8} 
%Mean of MSE                            &                     & 0.7707         & 0.5457           & 0.5413          & 2.8935           & 0.8974           & 0.8318             \\ \cline{1-1} \cline{3-8} 
%Variance of MSE                        &                     & 0.0000          & 0.0000          & 0.0000          & 0.0011            & 0.0002           & 0.0001           \\ \hline
\end{tabular}
\end{subtable}
\label{tbl:sim_mse_pca}
\end{table}

Table \ref{tbl:sim_mse_pca}(a) (respectively, Table \ref{tbl:sim_mse_pca}(b)) summarizes the results, each of which is based on 100 simulation replications, of a simulation study of the performance of 2SDR in the PCA model (respectively, the ${\rm H_MPCA}$ model). Note that if the actual ranks $p_0^*, q_0^*$ and $r^*$ are all known, 2SDR has computational complexity of the order $O(p^2p_0^*+q^2q_0^*)$+$O((p_0^*q_0^*)^2r^*)$, in which the first summand is that of MPCA, in contrast to $O(p^2q^2r)$ for PCA that vectorizes the image without going through MPCA first. Table \ref{tbl:sim_mse_pca}(a) assumes that the data are generated from the PCA model ~\eqref{eq:PCA_vec} with $y=\vec(X)$ and $pq \times r^*$ random matrix $\Gamma$ with orthonormal columns,
\begin{equation}\label{eq:setting}
p = q = 40,r^*=25; \mu = 0, c^{-1} \epsilon \sim N(0,I_{pq}), \nu \sim N(0, \Delta),   %\ \mbox{ or } T_5, %\footnote{$\nu$ is also generated form $N(0,I) \mbox{ or } T_5$.}
\end{equation}
where $\Delta = \DIAG (\delta_1,\delta_2,\cdots,\delta_{r^*})$ with $\delta_i = 10(26-i)$, and the value of $c$ is listed in Table \ref{tbl:sim_mse_pca}(a) alongside the sample size $n$ used to fit the model. Note that this choice of relatively small $p,q$ and $r^*$ makes fitting the PCA model computationally feasible. We apply the SURE criterion for fitting MPCA  model which is incompatible with the data generating mechanism~\eqref{eq:setting}, and we get $\hat p_0=\hat q_0=1$ in all 100 simulations. Although the SURE criterion fails to choose a reasonable rank, we still carry out MPCA and 2SDR by assuming that nominal values of $p_0=q_0=10$ in running  100 simulations of the (wrong) working model and the actual values $r^*=25$. However,
%To simplify the computational task in running  100 simulations for each setting in Table \ref{tbl:sim_mse_pca}, we assume that nominal values of $p_0=q_0=10$ of the (wrong) working model and the actual values $r=25$.  
%and $\sigma^2$ are estimated by the Random Matrix Theory in \cite{Iping2019}.
%{\color{red}$\sigma^2$ are known and used in implementing 2SDR so that they do not need to be estimated. Table 1 of \cite{Iping2019} shows that the estimate of $\sigma^2$ given there, as we have pointed out in the last paragraph of Section 1.3, "is quite accurate and the bias and standard error are below or around 1\% of $\sigma^2$"}.
 although 2SDR starts with the wrong working model when the data are generated from the PCA model~\eqref{eq:PCA_vec}, what really matters is the performance of image reconstruction, and fitting working models to a sample of size $n$  and dimension reduction are means to that end. A commonly used performance measure in image reconstruction is the Mean Squared Error (MSE)$= \sum_{i=1}^n\|\VEC(\hat{X_i}) - \vec({\rm true\  image})\|^2/(pqn)$, where $\hat X_i$ is the reconstructed image. Table \ref{tbl:sim_mse_pca}(a) gives the simulation results of MSE for 2SDR, in comparison with fitting the PCA model that generates the data and with fitting the wrong  MPCA model. The four settings in Table \ref{tbl:sim_mse_pca}(a) cover two sample sizes $(n=1000, 100)$ and two signal-to-noise ratios (SNR= 0.5, 0.1, corresponding to $c= 4, 20$) defined by  $E(||\Gamma \nu||^2)/(pqc)$. It shows that 2SDR has MSE comparable to PCA for $n=100$ and has about $1/2$ of the MSE of PCA for $(n,c)=(100,20)$.
Table \ref{tbl:sim_mse_pca}(b) assumes that the data are generated from the ${\rm H_MPCA}$ model with 
\vspace{-.05in}
\begin{eqnarray}\label{eq:setting2}
&&\\
&&p = q = 50,p_0^*=q_0^*=8, \sigma^{-1} \vec(\cE) \sim N(0,I_{pq}) \mbox{ for its MPCA component,} \notag \\ 
&&r^*=8 ,c^{-1} \epsilon \sim N(0,I_{p_0^*q_0^*}) \mbox{ and } c=1.001\sigma^2 \mbox{ for subsequent PCA component,} \nn 
\end{eqnarray}
in which $\kappa_i = 40(9-i), \mbox{ with } 1 \le i \le r^*$, $A$ (respectively, $B$) is a $p \times p_0^*$ (respectively, $q \times q_0^*$) matrix with orthonormal columns, which are the eigenvectors of the ordered eigenvalues of $(X-M)P_{B}(X-M)^{\top}$ (respectively, $(X-M)P_{A}(X-M)^{\top}$); see the second paragraph of Section \ref{MPCA_PCA}. The four settings in Table \ref{tbl:sim_mse_pca}(b) cover two sample sizes $(n=1000, 100)$ and two signal-to-noise ratios (SNR= 0.5 corresponding to $\sigma^2$= 1.1, 0.1 corresponding to $\sigma^2$= 5.6); here  SNR =$ \{\sum_{i=1}^{r^*}(\kappa_i-c)\}/(pq\sigma^2+p_0^*q_0^*c)$. %Figure \ref{fig:simu_mse} plots these MSE values accordingly.  
%The ranks $(\hat p_0, \hat q_0)$ are then estimated by the SURE criterion
% $\hat r_{\rm GIC}$ by GIC criterion and $\sigma^2$ are estimated by the Random Matrix Theory %\cite{Iping2019} in running 100 simulations.
%Later we carry out another simulation study reported in Table \ref{tbl:sim_rank} to show the accuracy rates of rank selection.
Table \ref{tbl:sim_mse_pca}(b) shows that
%even when the underlying model does not favor 2SDR, 2SDR outperforms PCA in setting~$4$, where both SNR and sample size are comparatively small. On the other hand, 
2SDR outperforms both MPCA and PCA in all settings, particularly for relatively small sample size $n=100$ or low SNR ratio $=0.1$. %The differences become more obvious when the sample size is small or when the SNR is low. In short, 2SDR is the best choice under mode ${\rm H_MPCA}$ that 2SDR  reaches the lowest MSE among the three.

%%%%%%%%%%%%%%%%
%% Table 2
%%%%%%%%%%%%%%%%
\begin{table}[!t]
\captionsetup{justification=centering, singlelinecheck=false}
\centering
%\captionsetup{justification=centering}\centering
\begin{threeparttable}
\caption{Accuracy rates of $\hat r_{\rm GIC}$, AIC and BIC}
\begin{tabular}{@{}c|c|c|c?c|c|c@{}} 
\toprule
                  & \multicolumn{3}{c?}{$\text{SNR}=0.5$} & \multicolumn{3}{c}{$\text{SNR}=0.1$}       \\ \toprule
                   & AIC   & BIC              & $\hat r_{\rm GIC}$     &        AIC   & BIC              & $\hat r_{\rm GIC}$                     \\ \midrule
Gaussian              & 0.96            & 1.00           & 1.00           & 0.98            & 1.00  & 1.00    \\ \midrule
$T_5$          &  0.00            & 0.85            & 0.97            &  0.00         & 0.82  & 0.93     \\  \bottomrule
\end{tabular}
\label{tbl:sim_rank}
\end{threeparttable}

\end{table}

We next conduct a simulation study of the accuracy ratio of $\hat p_0, \hat q_0$ using the SURE criterion under the ${\rm H_MPCA}$ model for $n=1000$ and $\sigma^2=1.1,5.6$, not only for the Gaussian model for $\sigma^{-1}\vec(\cE)$ and $c^{-1}\epsilon$ but also for the non-Gaussian case $\sigma^{-1}\vec(\cE) \sim T_5$ and $c^{-1}\epsilon \sim T_5$, where $T_5$ is the $pq$-dimensional (or $p^*_0q^*_0$-dimensional, for $c^{-1}\epsilon$) t-distribution with 5 degree of freedom. All of the 100 simulations give $(\hat p_0, \hat q_0)=(p^*_0, q^*_0)$. We then consider the accuracy rate ${\rm P}(\hat r_{\rm GIC}=r^*)$ in the PCA stage of 2SDR and compare it with AIC and BIC that assume $p^*_0, q^*_0$ to be known. The results are given in Table \ref{tbl:sim_rank} and show that $\hat r_{\rm GIC}$ has marked improvement over AIC and BIC in the non-Gaussian case of $T_5$. %accuracy rates of the SURE criterion~\eqref{first_stage} in the first stage as well as the accuracy rates of various criteria in the second stage including AIC, BIC, GIC and PCA's SURE \cite{ulfarsson2008dimension} upon 100 repetitions.
%\footnote{For PCA's SURE %method, the formula can be %obtained by setting %$q_0=q=1$ %in~\eqref{mpca_df}. In %addition, we use the %algorithm depicted in %\cite{ulfarsson2008dimensio%n} to estimate the variance %of noise in the SURE %criterion.} 
%The rank selection reaches $100\%$ accuracy in the first stage for both SNR equals to 0.5 or 0.1 and for both Gaussian and Student's t-distributions. In the second stage, all the information criteria, except AIC, reach 100$\%$ accuracy rate under Gaussian distribution. For the  Student's t-distribution, GIC performs better than the other criteria.

%%%%%%%%%%%%%%%%%%%%%%%%%%%%%%%%
%%         Section 3         %% 
%%%%%%%%%%%%%%%%%%%%%%%%%%%%%%%%
\section{Cryo-EM applications}
\label{c:application}
\label{c:sim}
%\subsection{Synthetic E. coli 70S ribosome dataset}\label{s:syn_cryoem} 
In this section we apply 2SDR to the analysis of cryo-EM benchmark datasets in Sections 
\ref{Evaluste2SDR}
and  
\ref{Relion_set}
 dealing with 70S ribosome and 80S ribosome.  A ribosome is made from complexes of RNAs  (ribonucleic acids) that are present in  all  living cells to perform protein synthesis by linking amino acids together in the order specified by the codons of mRNA (messenger RNA) molecules to form polypeptide chains. A 70S ribosome comprises of a large 50S
 subunit and small 30S subunit; the "S" stands for Svedberg, a unit of time equal to $10^{-13}$ seconds, measuring how fast molecules move in a centrifuge. Eukaryotic ribosomes are also known as 
 80S ribosomes and have a large 60S subunit and small 40S  subunit. 
 
In a cell or  a virus, biological processes are carried out by numerous nano-machines made of protein
complexes. When the structures of these protein machines are visualized, they go through a series of functionally relevant conformations along the time trajectory.
Because the protein sample for a cryo-EM experiment is under an almost physiological environment,
 it usually collects various conformations of the protein structure.
However, a homogeneous conformation dataset is a basic requirement to push  high resolution of the 3D density map. Thus,  3D clustering which aims to separate the data into subsets of  homogeneous conformation  has become an effective approach in cryo-EM data analysis \cite{katsevich2015covariance, penczek2011identifying, Serna2019}. In the past decade, the linear subspace model that represents the protein motion using the eigenvolumes from the covariance matrix of 3D structures is an active research area. This technique not only can analyze discrete heterogeneity \cite{katsevich2015covariance, penczek2011identifying}, but can also obtain energy landscape associated with the 3D structures \cite{anden2018structural,haselbach2017long, haselbach2018structure}. In all these approaches, PCA plays an important role to estimate the top eigenvolumes. However, volume vectorization   may produce enormous dimensionality and the traditional solution by voxel binning usually results in blurring of the variations among groups and degrading the clustering performance \cite{anden2018structural}. 
 Section \ref{subsec: variab}
 uses a heterogeneous dataset of 2D cryo-EM images to analyze the 3D variability of the reconstructed images. 
 
%%%%%%%%%%%%%%%%
%% Figure 2
%%%%%%%%%%%%%%%%
\begin{figure}[]
\centering
%%\subcaptionbox{\footnotesize{Eigenspace estimation (Higher is %%better)}}{\includegraphics[width=0.50\textwidth]{Figure/pca_3.jpg}}%
%%\hfill % <-- Seperation
\includegraphics[width=0.8\textwidth]{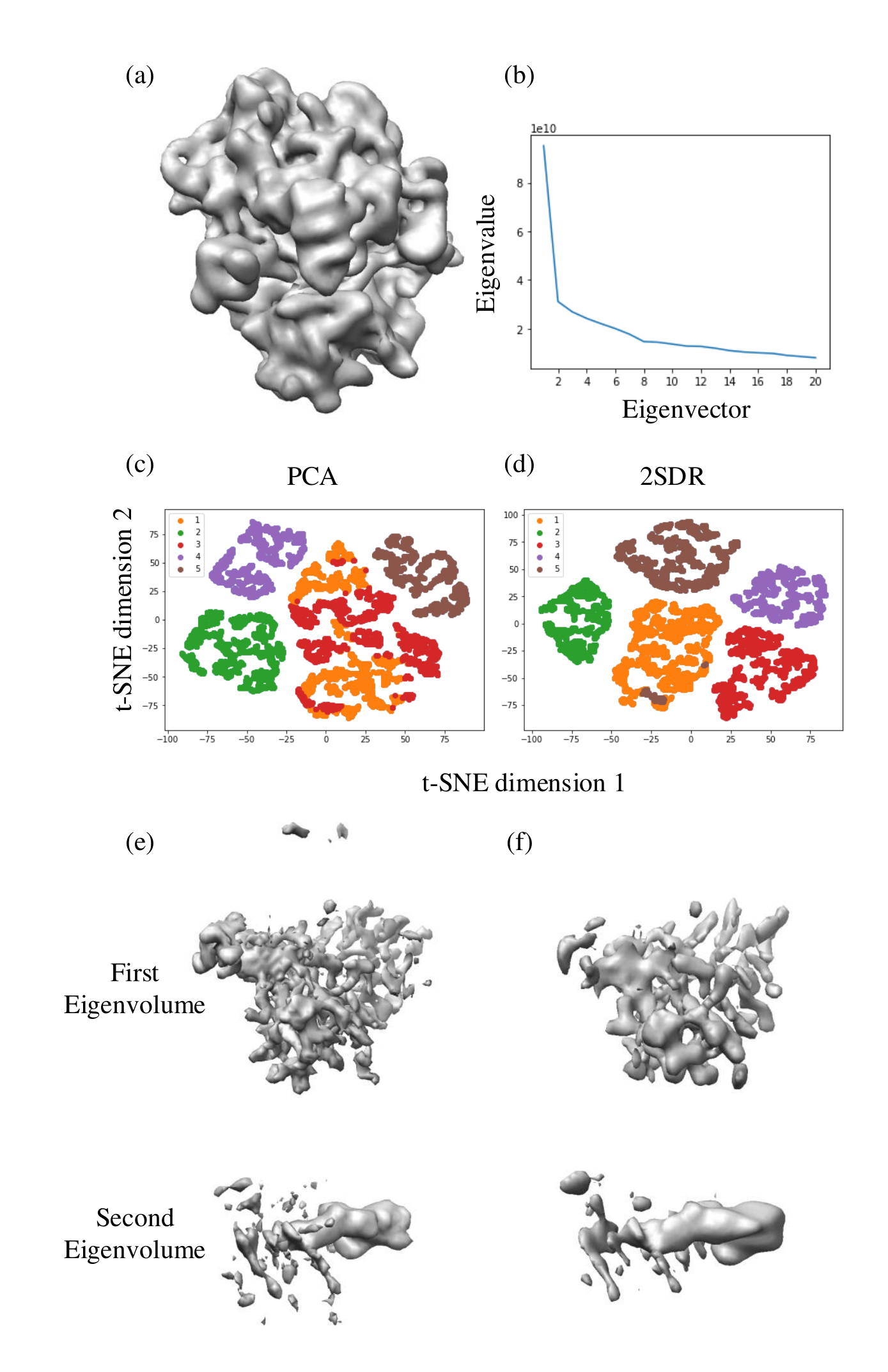}%
\caption{3D variability analysis. (a) The average volume.  (b) The Scree plot that shows leading 20 eigenvalues. (c)(d) The scatter plot of t-SNE embedding of 8 factorial coordinates computed by  PCA and 2SDR approaches and the color labels are according to the class assignments by k-means. (e) The first and second eigenvolume solved by performing PCA on 11,000 resampled volumes. (f) Same as (e) but with 2SDR.}
\label{fig:eig_volume}
\end{figure}

\begin{figure}
%\centering
%%\subcaptionbox{\footnotesize{Eigenspace estimation (Higher is %%better)}}{\includegraphics[width=0.50\textwidth]{Figure/pca_3.jpg}}%
%%\hfill % <-- Seperation
\subcaptionbox{\footnotesize{}}{\includegraphics[width=0.48\textwidth]{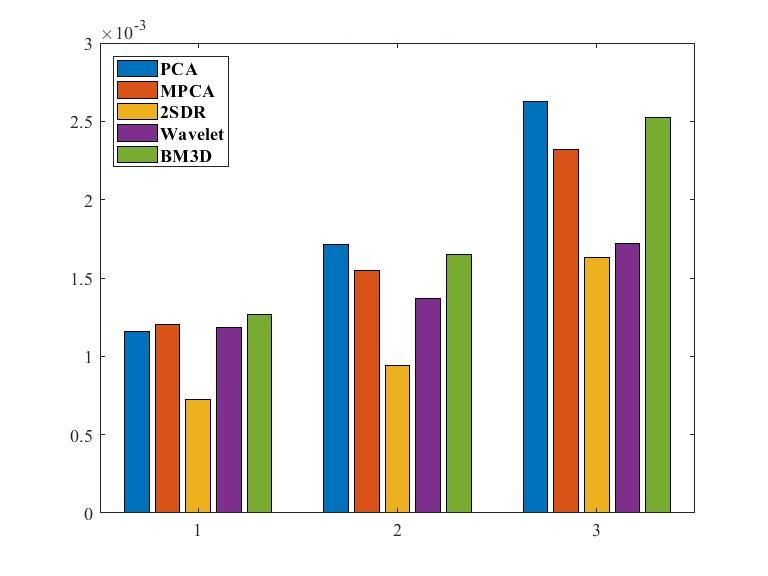}}%
%\hfill % <-- Seperation
\subcaptionbox{\footnotesize{}}{\includegraphics[width=0.48\textwidth]{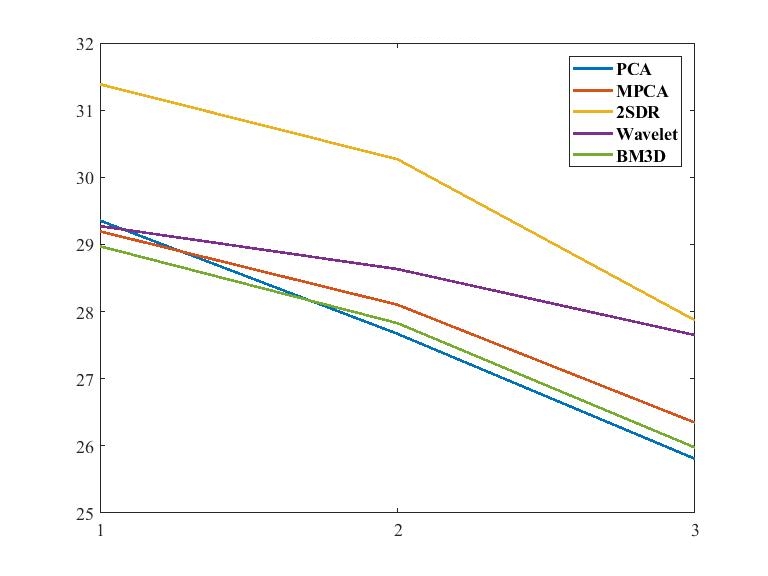}}
\caption{(a) MSE  and (b) PSNR of synthetic E. coli 70S ribosome dataset using PCA, MPCA, 2SDR, Wavelet and BM3D. Three simulations under different settings are tested: Label 1 for SNR $= 0.09$; Label 2 for SNR $=0.06$;  Label 3 for SNR $=0.03$. The details results of each setting are in the Table \ref{tbl:syn_mse}.}
\label{fig:syn_mse}
\end{figure}

\begin{table}[!b]
\centering
\captionsetup{justification=centering}
\caption{MSE and PSNR  comparison for various dimension reduction methods on synthetic E. coli 70S ribosome dataset.}  %$n = 5000, p=130, q=130$ in all settings. {The rank $(\hat p_0, \hat q_0, \hat r_{\rm GIC})$ selected here is  (15,15,53) for Setting 1; (14,14,48) for Setting 2; and (11,11,33) for Setting 3 .}}%according to our rank selection criterion.}}
\begin{tabular}{@{}c|c|c|c|c|c@{}}
\toprule
Method                      & PCA              & MPCA              & 2SDR              & Wavelet  &  BM3D              \\ \midrule
\multirow{1}{*}{Parameters} & \multicolumn{5}{c}{$\text{Setting}~1: \text{SNR}=0.09$}        \\ \toprule
 MSE  ($10^{-3}$)          & 1.1435            & 1.2040           & 0.7268           & 1.1825            & 1.2669                       \\ \midrule
Std of MSE ($10^{-5}$)      & 0.4056             & 0.0985            & 0.0490            & 0.0748             & 0.0501                       \\     \midrule
 PSNR                      & 29.4176            & 29.1939           & 31.3860           & 29.2718            & 28.9727                      \\ \midrule
Std of PSNR ($10^{-1}$)      & 0.1540             & 0.0356            & 0.0295            & 0.0274             & 0.0173                    \\     \midrule
\multirow{2}{*}{} & \multicolumn{5}{c}{$\text{Setting}~2: \text{SNR}=0.06$}        \\ \toprule
 MSE ($10^{-3}$)           & 1.7109           & 1.5484           & 0.9402           & 1.3703            & 1.6493                    \\ \midrule
Std of MSE ($10^{-5}$)     & 0.5474            & 0.1323            & 0.0742            & 0.1034             & 0.1609                  \\     \midrule
 PSNR                     & 27.6679            & 28.1010           & 30.2676           & 28.6319            & 27.8270                 \\ \midrule
Std of PSNR ($10^{-1}$)      & 0.1390             & 0.0371            & 0.0344            & 0.0328             & 0.0424                \\     \midrule
\multirow{2}{*}{}           & \multicolumn{5}{c}{$\text{Setting}~3: \text{SNR}=0.03$}      \\ \toprule
%Method                      & PCA              & MPCA              & 2SDR                \\ \midrule
 MSE  ($10^{-3}$)         & 2.6263            & 2.3199           & 1.6331           & 1.7182            & 2.5267                    \\ \midrule
Std of MSE($10^{-5}$)      &  0.5925            &  0.2107           & 0.1315            & 0.1616             & 0.2536                \\     \midrule
 PSNR                     & 25.8066            & 26.3453           & 27.8698   & 27.6493             & 25.9744             \\ \midrule
Std of PSNR ($10^{-1}$)     &  0.0980            &  0.0395           & 0.0349      & 0.0409               & 0.0436  \\ \bottomrule
\end{tabular}
\label{tbl:syn_mse}
\end{table}

\subsection{3D variability analysis using a heterogeneous  dataset of 2D images} \label{subsec: variab}

%Besides the low SNR nature of images and the huge computation workloads, the heterogeneity of cryo-EM dataset brings another challenge when analyzing the images. In contrast to X-ray crystallography that measures ensembles of particles, cryo-EM collects 2D images of individual particles. Since each cryo-EM image contains a unique instance of the molecule, the projection images of multiple conformations of the target macromolecule may co-exist within a  dataset. A direct approach to address the heterogeneity is through 3D clustering to classify various conformations of the molecule structures. 

We follow 
%Penczek and his colleagues 
\cite{penczek2011identifying} to generate a dataset containing 9,453 2D particle images projected from  five 70S ribosome conformations with minor differences 
due to combinations of the absence or presence of tRNA (transfer RNA) and EF-G (elongation factor G). Next, we resampled these  particle images to generate 11,000 3D volumes (density maps)  on $75 \times 75 \times 75$ voxels. 
We then solve the eigenvolumes using PCA or 2SDR,\footnote{ To perform 2SDR, the rank of each mode for the first stage is set to $80 \%$ of explain variance ratio; the rank for the second stage is 8 suggested by the elbow method of the scree plot.} and compare the performance of these two methods
using the factorial coordinates defined in \cite{penczek2011identifying}.

As shown in Figure \ref{fig:eig_volume}, more broken portions appear in the eigenvolumes solved by
PCA, which suggests that the eigenvolumes solved by 2SDR is more reliable. This is confirmed
by the t-SNE plots and k-means with 5 classes on their factorial coordinates. 
Since each particle image corresponds to a projection of one of the five conformations, we expect to see five clusters among the factorial coordinates of all the particle images.
The  t-SNE plot can separate 5 groups better for 2SDR approach than that for PCA as shown in Figure \ref{fig:eig_volume}. The clustering performance is also quantitatively evaluated through impurity and c-impurity \cite{schutze2008introduction} defined as follows. Let $\{c_i\}$ be sets of true class labels, $\{w_j\}$ be sets of predicted cluster labels, and $|.|$ be the cardinality of the set. The impurity and c-impurity are defied as 
\begin{align*}
{\rm impurity}= 1 - n^{-1}\sum_j\max_i|c_i \cap w_j| \\
\mbox{ c-impurity}= 1 - n^{-1}\sum_i\max_j|c_i \cap w_j|
\end{align*}
The impurity is 0 if each predicted cluster contains only members of a single class. The c-impurity is 0 if all members of a given class label are assigned to the same predicted cluster. In summary, small values of the impurity and c-impurity indicate better performance of the clustering results. The impurity and c-impurity numbers are 0.01 for 2SDR whereas those for PCA are 0.1995 and 0.2413, respectively, showing the superiority of 2SDR in dimension reduction for 3D density maps.

%%%%%%%%%%%%%%%%
%% Figure 4
%%%%%%%%%%%%%%%%
\begin{figure}[p]
\centering
    \includegraphics[width=4.2in]{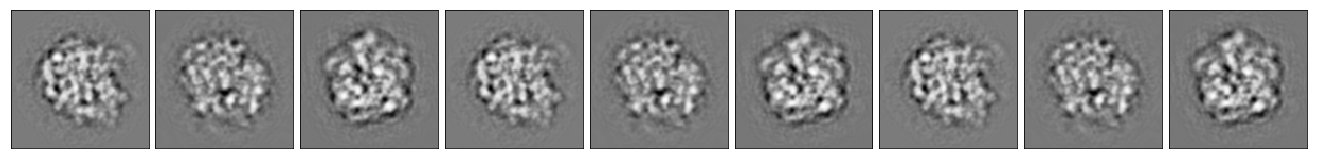}
      \includegraphics[width=4.2in]{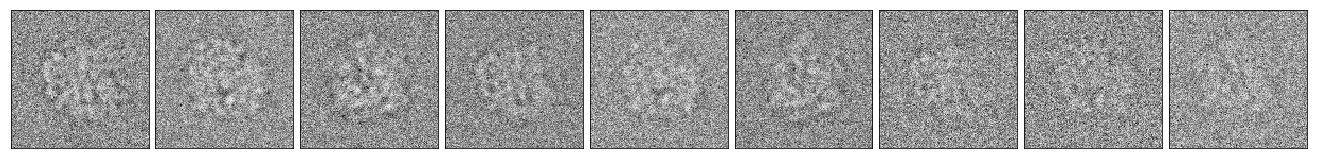}
        \includegraphics[width=4.2in]{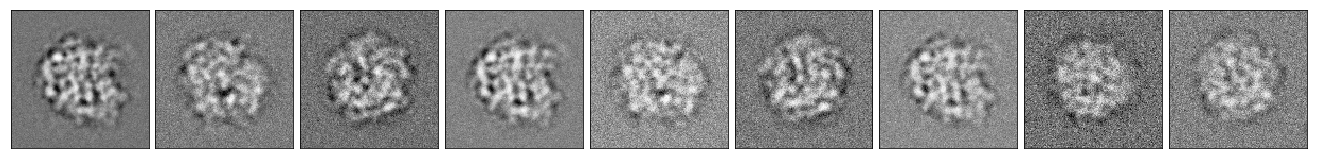}
          \includegraphics[width=4.2in]{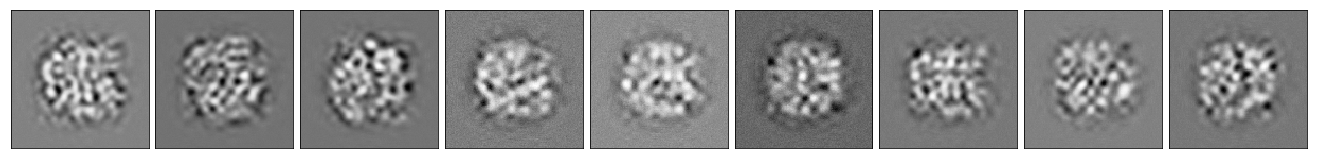}
          \includegraphics[width=4.2in]{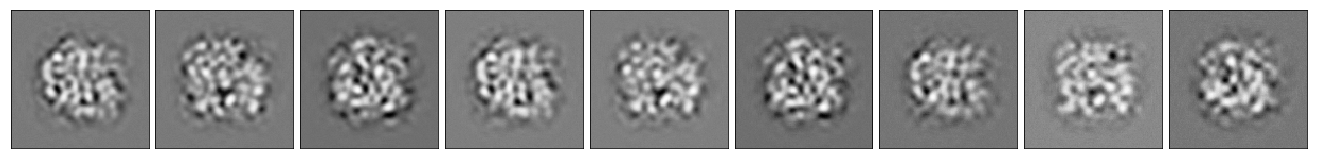}
           \includegraphics[width=4.2in]{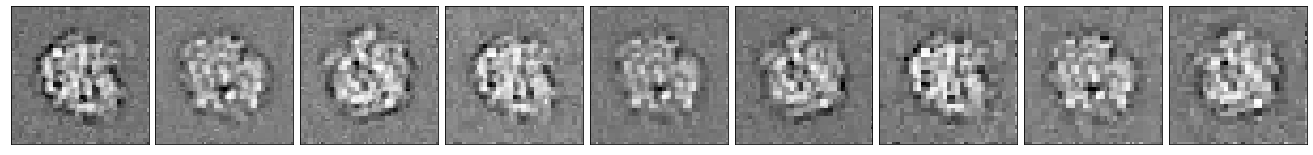}
            \includegraphics[width=4.2in]{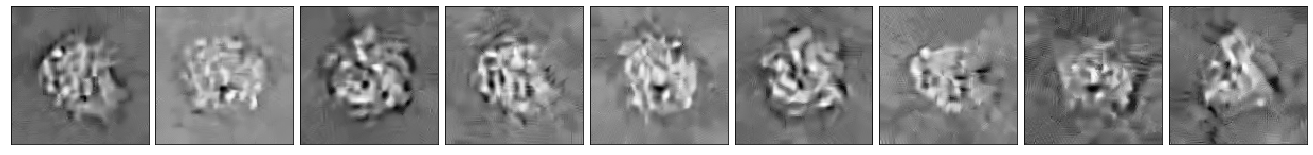}
\caption{Reconstruction of synthetic 70S ribosome images by 5 dimension reduction methods. The first row shows the 9 original clean synthetic images. The second row shows the noisy images corresponding to different SNR levels;  the first 3 columns from the left correspond to SNR$=0.09$, the middle 3 columns correspond to SNR$=0.06$ and the last 3 columns correspond to SNR$=0.03$. The third row is the result by applying PCA to the second row with $\hat r_{\rm GIC}=53$ (first 3 images from the left), $\hat r_{\rm GIC}=48$ (middle 3 images) and $\hat r_{\rm GIC}=33$ (last 3 images). The fourth row is the result by applying MPCA with $(\hat p_0, \hat q_0)=(15,15)$ for the first 3 images from the left, $(14,14)$ for the middle 3 images, and $(11,11)$ for the last 3 images. The fifth row is the result by applying 2SDR, with the same choice of  $(\hat p_0, \hat q_0)$ as in MPCA followed by that of $\hat r_{\rm GIC}= 53,~48,~33$. The sixth (respectively, seventh) row gives the result by applying Wavelet (respectively, BM3D). } 
\label{fig:synthetic_result}
%\end{figure}

%\begin{figure}[]
\centering
%%\subcaptionbox{\footnotesize{Eigenspace estimation (Higher is %%better)}}{\includegraphics[width=0.50\textwidth]{Figure/pca_3.jpg}}%
%%\hfill % <-- Seperation
\includegraphics[width=1\textwidth]{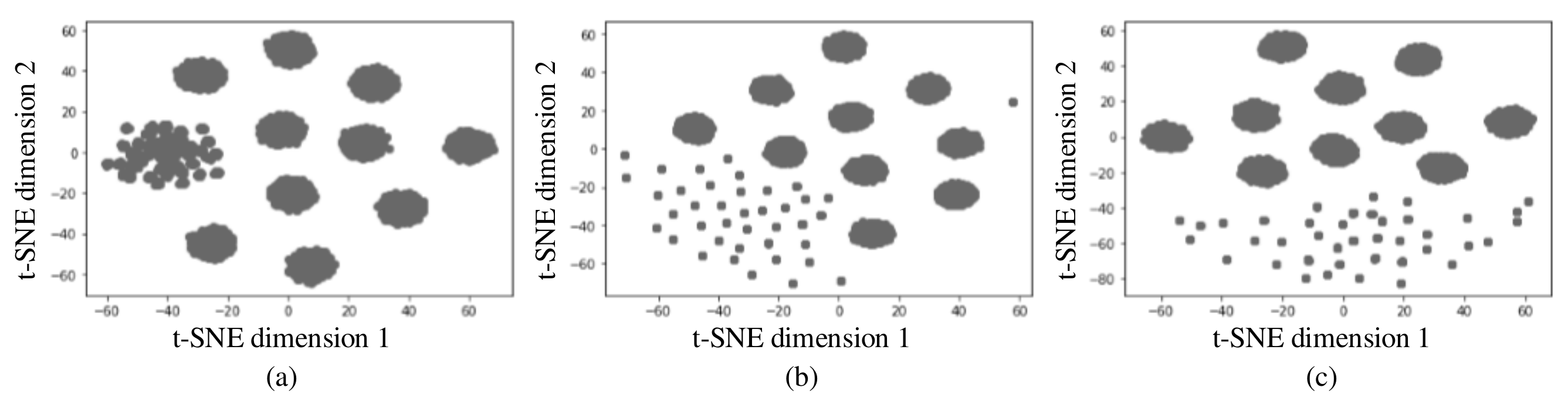}%
\caption{t-SNE embedding of (a) PCA,  (b) MPCA and (c) 2SDR at SNR$=0.03$ on synthetic E. coli 70S ribosome dataset.}
\label{fig:syn_tsne}
\end{figure}

%%%%%%%%%%%%%%%%
%% Figure 6
%%%%%%%%%%%%%%%%
\begin{figure}[]
\centering
    \includegraphics[width=4.8in]{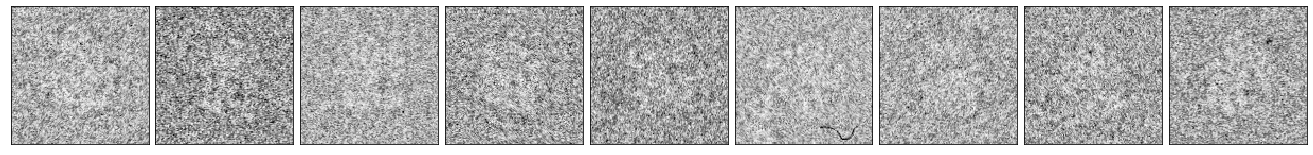}
      \includegraphics[width=4.8in]{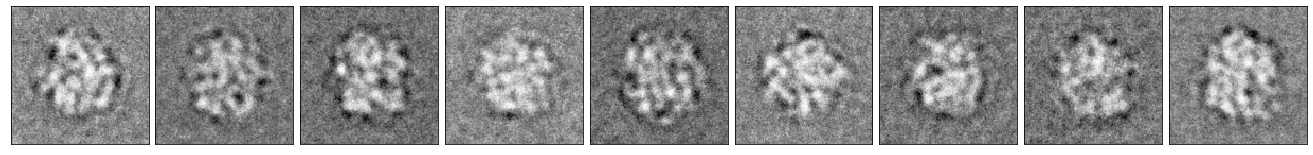}
        \includegraphics[width=4.8in]{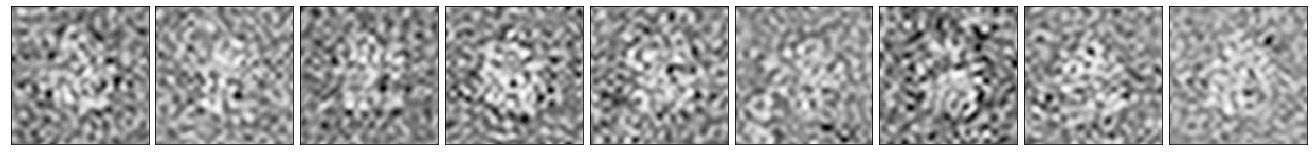}
          \includegraphics[width=4.8in]{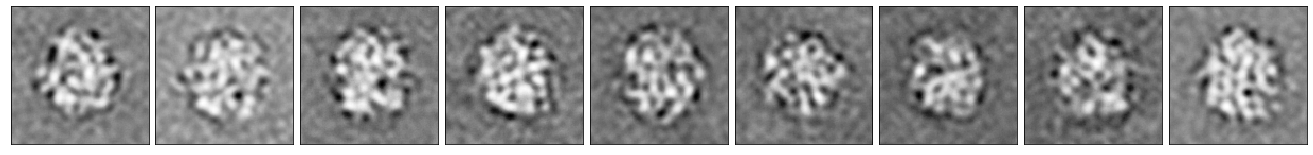}
           \includegraphics[width=4.8in]{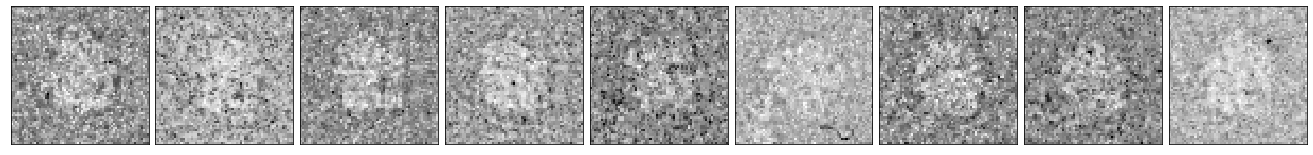}
            \includegraphics[width=4.8in]{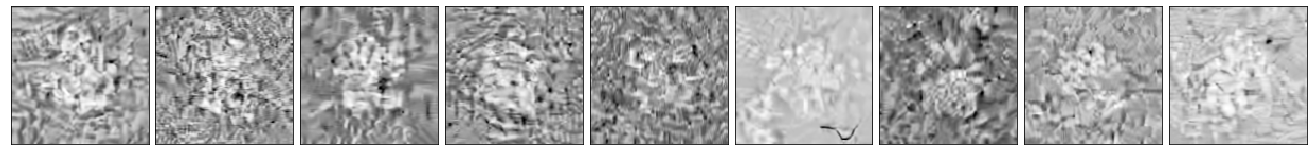}
\caption{Reconstruction of experiment E. coli 70S ribosome particle images by 5 dimension reduction methods. The first row shows the 9 original particle images. {The second row is by PCA with $\hat r_{\rm GIC}=60$ components. The third row is by MPCA with $(\hat p_0, \hat q_0)=(25,25)$, which contributes 625 basis components. The fourth row is by 2SDR: $(\hat p_0,\hat q_0)=(25,25)$ for MPCA followed by choosing $\hat r_{\rm GIC}=60$ for PCA. The fifth row is by Wavelet denoising and the sixth row is by BM3D.}}
\label{fig:ribsom70s}
\end{figure}

\subsection{Performance measure and applications to 70S ribosome data}\label{Evaluste2SDR}
Here, we apply our algorithm and the rank selection procedure on the synthetic cryo-EM data in this subsection where the dataset is prepared as follows.  We first downloaded the Relion \cite{scheres2012relion} classification benchmark dataset, \href{https://www.ebi.ac.uk/pdbe/emdb/test_data.html}{E. coli 70S ribosome}, which  contains 10000 particle images with box size  $130 \times 130$. The first 5000 and the second 5000 images of this dataset represent different structure conformations. Second, we apply CryoSparc \cite{punjani2017cryosparc} to generate  the 3D density map from the first 5000 images referred to the ribosome bound with an elongation factor (EF-G). Then, a total of 50 distinct 2D images with $130 \times 130$ pixels were generated by projecting the 3D density map in equally spaced (angle-wise) orientations. Third, 5000 images were generated from these 50 projections.\footnote{Here, to reflect the fact that real data is often collected with preferred orientations, 10 projections are repeated with 400 copies and the other projections are repeated with 25 copies.} Each image was then convoluted with the electron microscopy contrast transfer function randomly sampled from a set of 50 CTF values.\footnote{Here, the defocus is randomly sampled from 2.1um to 3.5um and the astigmatism angle is from 0.2 to 1.4 radian. The electron beam accelerating voltage was set to be 300KeV with spherical aberration Cs = 2mm, amplitude contrast=0.07 and pixel size is 2.82\si{\angstrom}.} Finally, i.i.d. Gaussian noise $N\left(0,\sigma^2I_{pq}\right)$ with different $\sigma^2$ is added to generate 3 datasets such that the SNR is equal to 0.09, 0.06 and 0.03, respectively. %Since the true rank is not known in this case, we use the MSE between original clean data and reconstruction data to evaluate the rank selection performance. We  observe that BIC and GIC perform better than the other criteria upon 100 repetitions for different SNR levels as shown in Table \ref{tbl:syn_rank}. %in Section \ref{s:rank_selection}.

Quantitative comparison on five dimension reduction or denoising methods including   PCA, MPCA, 2SDR,  Wavelet \cite{chang2000adaptive} and BM3D \cite{dabov2007image} are then conducted on the synthetic dataset.  
 MSE and Peak Signal to Noise Ratio (PSNR) are computed based on 100 replica simulation with three SNR levels, where PSNR is defined as 
\begin{equation}\label{eq:psnr}
{\rm PSNR}(X_i, \hat X_i) = 10 \times \log_{10}\frac{{\rm Range}(X_i)^2}{{\rm MSE}},
\end{equation}
where $ {\rm Range}(X_i)$ is the value range according to the datatype of $X_i$.\footnote{For instance, if the data type of image is unit8 then it is 255.}
%As  SNR goes low, 2SDR has comparatively robust MSE  while  both PCA and MPCA increase their MSE significantly. 
MSE and PSNR in Figure \ref{fig:syn_mse} show that 2SDR outperforms all the other methods. The images reconstructed by these methods are presented in Figure \ref{fig:synthetic_result}, where the SNR is decreased from left-hand side to right-hand side. When SNR is 0.09, all methods can reconstruct the particles well.  As the noise increases, Figure \ref{fig:synthetic_result} shows that the particles reconstructed by 2SDR and PCA match the original images much better than those by MPCA, Wavelet and BM3D. 
When SNR drops to 0.03, 2SDR performs better than PCA in regarding to the contrast and background noise reduction. 
 %As SNR is 0.01, MPCA reconstructs square noisy particles which totally corrupt the particle information. 
 We further apply  t-SNE on the scores solved by PCA, MPCA and 2SDR at SNR$=0.03$ in Figure \ref{fig:syn_tsne} and observe that MPCA and 2SDR can perfectly separate the 50 clusters of the synthetic dataset while PCA tends to have small groups aggregate together. Overall speaking,  2SDR performs the best and can prepare the data best for clustering.

%While it can save lots of computation time using 2SDR, we would like to demonstrate that it possible to maintain competitive performance as PCA. To this end, the 
After the encouraging results obtained from the synthetic dataset, we proceed to test 2SDR on experiment dataset. The first 5000 particle images from the 70S ribosome dataset are used for demonstration, where Sparx \cite{hohn2007sparx} package for alignment. To implement the  rank selection procedure introduced in Section 2.1,  we choose $(p_u, q_u)$ as the $35\%$ of total variance on both column and row spaces. We  present the reconstructed images of 9 randomly selected particles by PCA, MPCA, 2SDR, Wavelet and BM3D in Figure \ref{fig:ribsom70s}.  MPCA, Wavelet and BM3D do not perform well in presenting the particle shapes on this real dataset.

%%%%%%%%%%%%%%%%
%% Figure 7
%%%%%%%%%%%%%%%%
\begin{figure}[]
\centering
%%\subcaptionbox{\footnotesize{Eigenspace estimation (Higher is %%better)}}{\includegraphics[width=0.50\textwidth]{Figure/pca_3.jpg}}%
%%\hfill % <-- Seperation
\includegraphics[width=1\textwidth]{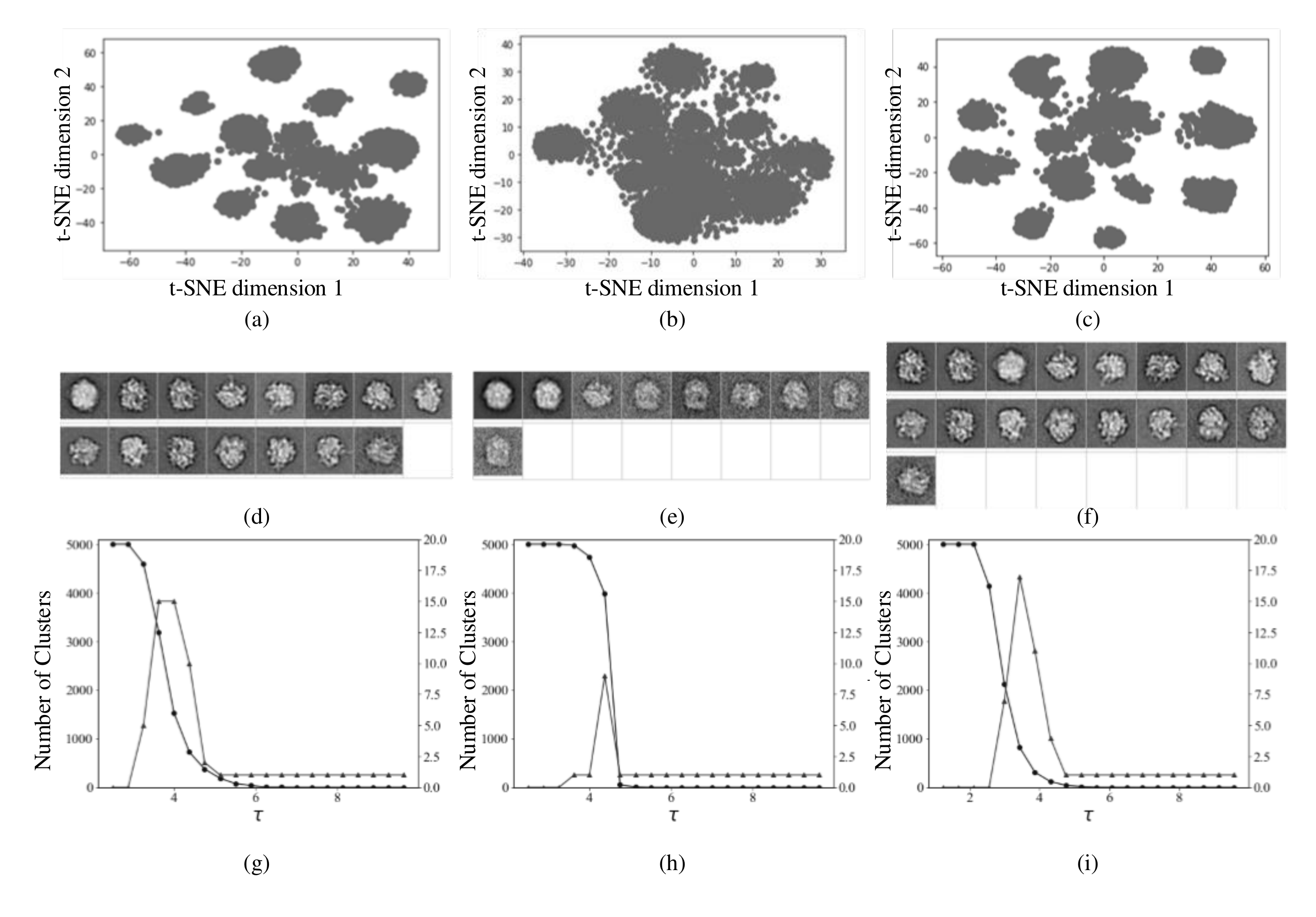}%
\caption{(a),(b),(c) from left to right, contains three t-SNE plots of the scores generated by PCA, MPCA and 2SDR; (d),(e),(f) presents the class averages  output by the clustering algorithm $\gamma$-SUP of PCA, MPCA and 2SDR; (g),(h),(i) shows the transition plot of $\gamma$-SUP on the experiment E. coli 70S ribosome dataset. %Note that there are two scale bar in 
In (g),(h),(i), the left scale bar corresponds to circle marker, which is the number of clusters, and the right scale bar corresponds to triangle marker, which is the number of clusters with cluster size larger than 10.}
\label{fig:tsne_70s}
\end{figure}

We apply  t-SNE on the scores solved by PCA, MPCA and 2SDR.\footnote{Here, the images are aligned using multireference alignment \cite{hohn2007sparx} for better visualization.} Figure \ref{fig:tsne_70s} shows that PCA and 2SDR can better separate the groups representing different orientated projections. The $\gamma$-SUP is further applied on the scores to evaluate the performance. Since $\gamma$-SUP is insensitive to the parameter $s$, we set it to a small value $s=0.025$ as suggested in the original paper \cite{AASChen2014}. The scale parameter $\tau$ is selected using phase transition plot as follows. We first normalize the scores and set an upper bound $\tau$ such that $\gamma$-SUP groups all points into one cluster. Second, we recursively divide the $\tau$ by 2 until each points forms one cluster and record it as the lower bound. Finally, we perform the grid search between the upper and lower bound and select the $\tau$ that maximizes the cluster number with the condition that each cluster is larger than a prescribed size. $\gamma$-SUP shows that 2SDR can produce the largest number of good classes and  the class averages from PCA and 2SDR show more structural details in Figure \ref{fig:tsne_70s} .

%%%%%%%%%%%%%%%%
%% Figure 8
%%%%%%%%%%%%%%%%
\begin{figure}[]
\centering
    \includegraphics[width=4.8in]{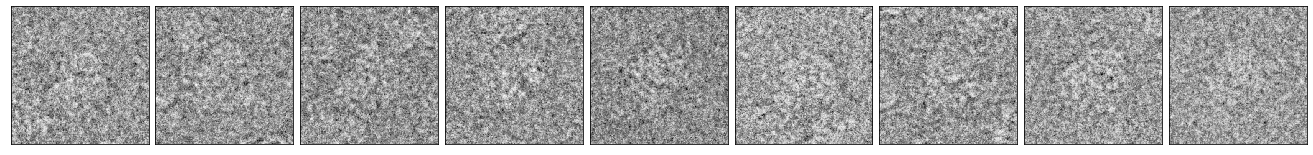}
        \includegraphics[width=4.8in]{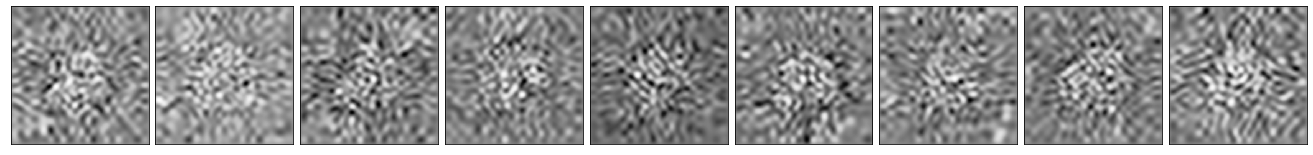}
          \includegraphics[width=4.8in]{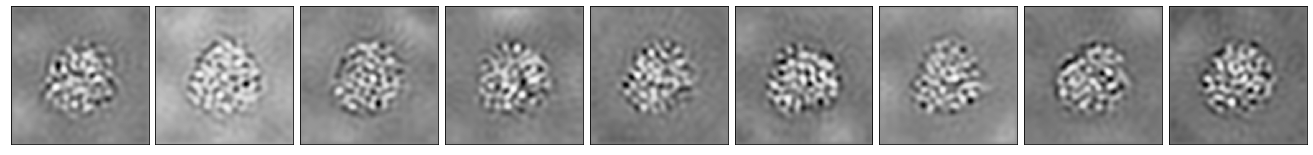}
           \includegraphics[width=4.8in]{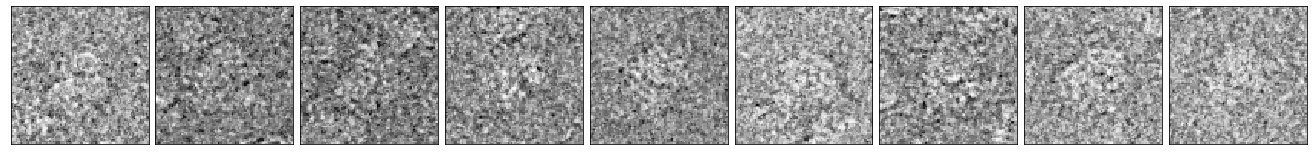}
            \includegraphics[width=4.8in]{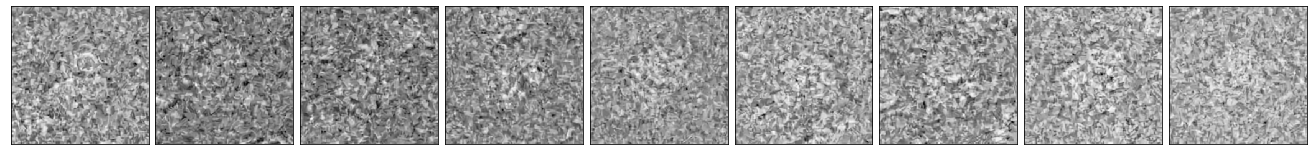}
\caption{Reconstruction of 9 randomly selected 80S ribosome particle images: 
%by MPCA, 2SDR, Wavelet and BM3D. The 
first row shows the original images,  second row the MPCA reconstruction with 
$(\hat p_0,\hat q_0)=(26,26)$, third row the 2SDR reconstruction with 
$(\hat p_0,\hat q_0)=(26,26)$ and $\hat r_{\rm GIC}=152$, fourth row the Wavelet reconstruction, and fifth row the 
BM3D reconstruction.}
\label{fig:ribsom80S}
\end{figure}

\subsection{Relion 80S ribosome benchmark dataset}\label{Relion_set}
To further demonstrate the computational advantage of 2SDR on large dataset, we test 80S ribosome  that comes from \href{https://www.ebi.ac.uk/pdbe/emdb/empiar/entry/10028/
} {Relion Benchmark example}. This dataset contains 105,247 particle images with pixel size 360 by 360 that many current PCA implementations fail to solve the complete set of eigenvectors due to the limitation of the underlying numerical linear algebra package LAPACK \cite{laug}.\footnote{ We used a server equipped with two Intel Xeon CPU E5-2699 v4 at 2.20GHz and 512GB memory to execute PCA for this dataset.}  In contrast, we can perform 2SDR in the server since the computational complexity has been reduced by several orders of magnitude. 
%https://stackoverflow.com/questions/49626572/computation-cannot-be-performed-with-standard-32-bit-lapack
%Though we can not obtain the full eigenvector set, top eigenvectors can still be extracted using algorithms like power method.} 
Figure \ref{fig:ribsom80S} shows nine randomly selected images and their reconstructions by  MPCA, 2SDR, Wavelet and BM3D; 2SDR clearly performs
much better than the other methods. 
\section{Discussion}
\label{c:discussion}
PCA was introduced in the early development of single particle cryo-EM analysis to reduce the dimension of the 2D projection images to facilitate 2D image clustering \cite{frank2006three, van1981use}. 
Recently Amit Singer and his coauthors \cite{bhamre2016denoising,zhao2016fast, zhao2014rotationally} introduced a method called "steerable PCA"
to deal with the random orientation nature of 2D projection images
by including all possible rotated 2D projections into the data covariance matrix to yield promising outcomes on particle denoising.  PCA has more general applications in cryo-EM analysis \cite{afanasyev2017single,van2016multivariate} such as analyzing discrete heterogeneity \cite{katsevich2015covariance,penczek2011identifying} and obtaining energy landscape associations with 3D structures \cite{anden2018structural, haselbach2017long,haselbach2018structure}. A disadvantage of applying PCA to these tensor structure data  lies in its computational bottleneck.

Many current statistical algorithms are designed for vector data. To handle matrix data or high-order tensor data, a naive approach is to accommodate the input format requirement by data vectorization, which can have prohibitive  computational cost. 
We have shown that 2SDR can overcome this computational difficulty and yet
have superior performance over PCA.
In addition, ${\rm H_MPCA}$ can be readily  extended to a mixture model of the form
\begin{eqnarray}
X \mid z &=& M + A  (U_z) B^\top +\cE\notag  \\
\vec(U_z)&=& \mu_z +G \nu+\epsilon,~~z\in\{1,\dots,K\}, \label{mixtureModel}\nn
\end{eqnarray}
where we introduce a random variable $z$ for cluster label and $\mu_z$ is the mean of $\vec(U_z)$,  $P(z=k)=\pi_k,~ k\in \{1,...,K\}$ and $\sum_{k=1}^K\pi_k=1.$
We have recently applied this extension to cryo-EM image analysis
and found excellent performance, and the promising results have led to further
development of this idea as our ongoing project. 
Moreover, 2SDR is not limited to the 2D cryo-EM image analysis.
We have also demonstrated 2SDR provides a promising alternative to PCA with better performance and
less computational overhead in the 3D heterogeneous volume analysis.
 In conclusion, 2SDR is a powerful innovation that can provide
 an alternative to PCA for dimension reduction in the analysis of
 noisy high-dimensional data.

\section*{Acknowledgement}
This work is supported by Academia Sinica:AS-GCS-108-08 and  MOST:106-2118-M-001-001 -MY2.

\bibliography{reference}

\clearpage

\end{document}